\documentclass{aa}
\usepackage{threeparttable,hhline,longtable}
\usepackage{deluxetable} 
\usepackage{epsfig,latexsym,graphicx,subfigure}
\usepackage{verbatim,amsmath}
\usepackage[T1]{fontenc}
\usepackage{ae,aecompl}
\usepackage{lscape}
\usepackage{natbib}
\newcommand{\ew}{pseudo-{\sc ew}}
\newcommand{\taa}{91T-like}
\newcommand{\sne}{SNe~Ia}
\newcommand{\sn}{SN~Ia}

\newcommand{\kmps}{km s$^{-1}$}
\newcommand{\dmft}{$\Delta m_{15}$}           
 
\begin{document}
 \title{Quantitative comparison between Type Ia supernova spectra at low and high redshifts: A case study}

\author{G.~Garavini\inst{1,4}\and
G.~Folatelli\inst{2}\and
S.~Nobili\inst{1}\and
G.~Aldering\inst{3}\and
R.~Amanullah\inst{1}\and
P.~Antilogus\inst{4}\and
P.~Astier\inst{4}\and
G.~Blanc\inst{5}\and
J.~Bronder\inst{6}\and
M.~S.~Burns\inst{7}\and
A.~Conley\inst{3,8}\and
S.~E.~Deustua\inst{9}\and
M.~Doi\inst{10}\and
S.~Fabbro\inst{11}\and
V.~Fadeyev\inst{3}\and
R.~Gibbons\inst{12}\and
G.~Goldhaber\inst{3,8}\and
A.~Goobar\inst{1}\and
D.~E.~Groom\inst{3}\and
I.~Hook\inst{6}\and
D.~A.~Howell\inst{13}\and
N.~Kashikawa\inst{14}\and
A.~G.~Kim\inst{3}\and
M.~Kowalski\inst{3}\and
N.~Kuznetsova\inst{3}\and
B.~C.~Lee\inst{3}\and
C.~Lidman\inst{15}\and
J.~Mendez\inst{16,17}\and
T.~Morokuma\inst{10}\and
K.~Motohara\inst{10}\and
P.~E.~Nugent\inst{3}\and
R.~Pain\inst{4}\and
S.~Perlmutter\inst{3,8}\and
R.~Quimby\inst{3}\and
J.~Raux\inst{4}\and
N.~Regnault\inst{4}\and
P.~Ruiz-Lapuente\inst{17}\and
G.~Sainton\inst{4}\and
K.~Schahmaneche\inst{4}\and
E.~Smith\inst{12}\and
A.~L.~Spadafora\inst{3}\and
V.~Stanishev\inst{1}\and
R.~C.~Thomas\inst{3}\and
N.~A.~Walton\inst{18}\and
L.~Wang\inst{3}\and
W.~M.~Wood-Vasey\inst{3,8}\and
N.~Yasuda\inst{19}
\vspace{2mm}\\(The Supernova Cosmology Project)}

\institute{Department of Physics, Stockholm University,  Albanova University Center, S-106 91 Stockholm, Sweden
\and Observatories of the Carnegie Institution of  Washington,  813 Santa Barbara St., Pasadena,  CA 91101 
\and E. O. Lawrence Berkeley National Laboratory, 1 Cyclotron Rd., Berkeley, CA 94720, USA 
\and LPNHE, CNRS-IN2P3, University of Paris VI \& VII, Paris, France 
\and Osservatorio Astronomico di Padova, INAF, vicolo dell'Osservatorio 5, 35122 Padova, Italy
\and Department of Physics, University of Oxford, Nuclear \& Astrophysics Laboratory,  Keble Road, Oxford, OX1 3RH, UK
\and Colorado College, 14 East Cache La Poudre St., Colorado Springs, CO 80903
\and Department of Physics, University of California Berkeley, Berkeley, 94720-7300 CA, USA
\and American Astronomical Society,  2000 Florida Ave, NW, Suite 400, Washington, DC, 20009 USA.
\and Institute of Astronomy, School of Science, University of Tokyo, Mitaka, Tokyo, 181-0015, Japan
\and CENTRA e Dep. de Fisica, IST, Univ. Tecnica de Lisboa
\and Department of Physics and Astronomy, Vanderbilt University, Nashville, TN 37240, USA
\and Department of Astronomy and Astrophysics, University of Toronto, 60 St. George St., Toronto, Ontario M5S 3H8, Canada
\and National Astronomical Observatory, Mitaka, Tokyo 181-0058, Japan
\and European Southern Observatory, Alonso de Cordova 3107, Vitacura, Casilla 19001, Santiago 19, Chile 
\and Isaac Newton Group, Apartado de Correos 321, 38780 Santa Cruz de La Palma, Islas Canarias, Spain
\and Department of Astronomy, University of Barcelona, Barcelona, Spain 
\and Institute of Astronomy, Madingley Road, Cambridge CB3 0HA, UK 
\and Institute for Cosmic Ray Research, University of Tokyo, Kashiwa, 277 8582 Japan}

 \offprints{G.~Garavini , gabri@physto.se}

 \date{Received ...; accepted ...}
 \authorrunning{G.~Garavini}

\titlerunning{Search for Spectral Evolution in high-redshift supernovae}

\date{} \abstract{ We develop a method to measure the strength of the
  absorption features in Type Ia supernova (SN Ia) spectra and use it
  to make a quantitative comparison between the spectra of Type Ia
  supernovae at low and high redshifts. In this case study, we apply
  the method to 12 high-redshift (0.212 $\leq$ z $\leq$
  0.912) SNe~Ia observed by the Supernova Cosmology Project . Through measurements of the strengths of these features and
  of the blueshift of the absorption minimum in Ca~{\sc ii}~H\&K, we show that the spectra of the
  high-redshift SNe~Ia are quantitatively similar to spectra of nearby
  SNe~Ia (z < 0.15).  One supernova in our high redshift sample, SN
  2002fd at z=0.279, is found to have spectral characteristics that
  are associated with peculiar SN~1991T/SN~1999aa-like supernovae.}

\maketitle

\section{Introduction}

Type~Ia supernovae (SNe~Ia) are excellent distance indicators and have
been used to show that the expansion of the Universe is currently
accelerating
\citep{1998Natur.391...51P,1998ApJ...493L..53G,1998ApJ...507...46S,1998AJ....116.1009R,1999ApJ...517..565P,
  2003ApJ...598..102K,2003ApJ...594....1T,2004ApJ...602..571B,2004ApJ...607..665R,2005AJ....130.2453K,2006A&A...447...31A,2007astro.ph..1041W}.

Recently, both the SNLS\footnote{http://www.cfht.hawaii.edu/SNLS/} and
ESSENCE\footnote{http://www.ctio.noao.edu/essence/} projects, which
aim to use hundreds of SNe~Ia to constrain the nature of dark energy through
the measurement of the equation-of-state parameter, have reported
their first results.
The control of the systematic uncertainties is critical to their
success. Among the possible systematic effects, evolution of the SNe
Ia population over cosmological time-scales is one of the most
important and least understood.

Spectra of SNe~Ia are well-suited to study potential evolutionary
effects.  For example, the average metallicity of the Universe
increases with cosmic time, so it is reasonable to expect that
high-redshift SNe~Ia are in environments that have lower average
metallicity than those of nearby SNe~Ia.  The effect on the spectral
energy distribution of a lower metallicity progenitor has been modeled
by \citet{1998ApJ...495..617H} and \citet{2000ApJ...530..966L}. These
studies find that such SNe~Ia, especially at early epochs, are
expected to show enhanced flux in the UV, weaker absorption features
in the optical and a shift in the absorption minima of optical
features to longer wavelengths.

With the large number of well-observed low-redshift supernovae now
available, a wide range of spectral diversity is being found (\citet[see e.g. ]{2006PASP..118..560B}). The physical origin of these
differences is still not completely understood making it difficult to
predict their possible evolution with redshift. Statistical studies
are useful to probe differences between high and low-redshift SN~Ia
data sets.  So far, few distant SN~Ia spectra have been compared
with low-redshift data sets in a quantitative manner
\citep{1998Natur.391...51P,2000ApJ...544L.111C,2004ApJ...602..571B,2003astro.ph..8185R,2003ApJ...589..693B,2005AJ....129.2352M,2005AJ....130.2788H,2006A&A...445..387B,2005ApJ...634.1190H,2006AJ....131.1648B},
and few spectroscopically confirmed high-redshift SNe~Ia have been
reported as peculiar (two SN~1991T/SN~1999aa-like SNe~Ia in
\citet{2005AJ....129.2352M}; one, as we will see, in the present
paper). Hereafter, we follow the convention of describing
SN~1991T/SN~1999aa-like SNe~Ia by using "\taa" to represent both SN~Ia
subtypes.

During 2000, 2001 and 2002, the Supernova Cosmology Project (SCP)
obtained spectra of 20 high-redshift SNe~Ia with FORS2 on the ESO
Very Large Telescope \citep{2005A&A...430..843L}. In this paper, we
analyze the 14 spectra  of 12 \sn\ with the highest signal-to-noise ratios and
perform a quantitative comparison between these spectra and the
spectra of low-redshift SNe~Ia. The definition of a newly introduced
spectral indicator for Type Ia supernovae, namely pseudo-equivalent
width, is given in section \ref{ew}. The data-sets we apply the method
to are presented in section \ref{data}.  The properties of the
pseudo-equivalent width for SNe~Ia are given in section \ref{results}
together with the result of the comparison between high- and low-redshift SNe~Ia.  Summary and conclusions are given in section
\ref{conc}.


\section{Method}
\label{ew}

We use two spectral indicators to compare high- and low-redshift SN~Ia
spectra:

1) the wavelength of the absorption minimum at $\sim$ 3750 \AA, which is largely due to Ca~{\sc ii}~H\&K, and

2) newly-introduced pseudo-EW measurements of features associated with Ca~{\sc
  ii}~H\&K, Mg~{\sc ii} and Fe~{\sc ii}.

\subsection{Pseudo-Equivalent Widths of \sne }

The spectra of Type Ia supernovae are very characteristic and, in
comparison with supernovae of other types, relatively
homogeneous. Spectral features are broad, reflecting the high
velocities of the ejecta ($\sim$10000 \kmps), and evolve with  phase. However,
differences between different Type Ia supernovae have been noted (see
\citet{1997ARA&A..35..309F} for a review). In some cases the
differences are dramatic and have resulted in the definition of SNe~Ia
sub-types.  

The pseudo-equivalent width (\ew), first described in
\citet{gastonthesis}, can be used as a spectral indicator. It is
described in detail in the following section.

\subsection{Definition}

The equivalent width, as used in stellar spectroscopy, can be used to
measure the strength of absorption features in supernova
spectra. However, the relationship between this quantity and the
physical properties of the SN~Ia ejecta is complex. In a SN~Ia
spectrum the
overlap of thousands of lines give rise to a "pseudo" continuum
thus a real continuum can not be identified.
Because of this distinction with the definition of  equivalent widths  we use the term "pseudo-equivalent widths " to
refer to our measurements.
This does not
prevent us from using this well-defined quantity, nor does it not
allow up from building a consistent set of measurements for all SNe~Ia
in our data-sets.

\begin{figure}[htb]
  \centering
  \includegraphics[width=\hsize]{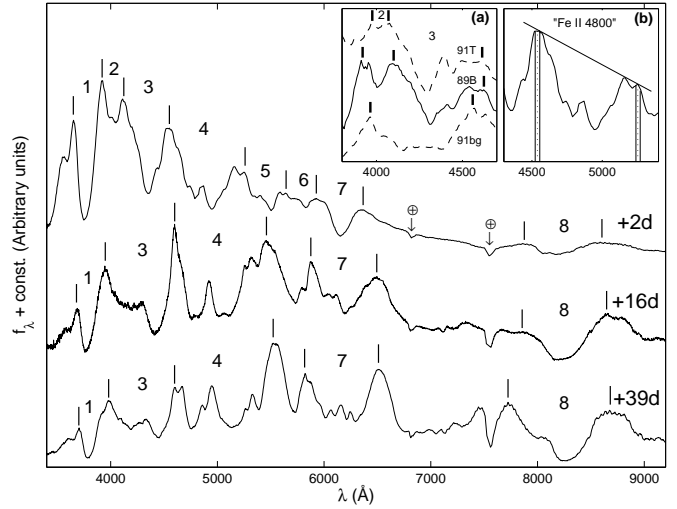}
  \caption{SNe~Ia spectral evolution and feature definitions for three
    epochs: 2, 16 and 39 days after maximum light. Numerical labels
    correspond to the following adopted feature names: {\bf 1}-
    ``Ca~{\sc ii} H\&K''; {\bf 2}- ``Si~{\sc ii} 4000''; {\bf 3}-
    ``Mg~{\sc ii} 4300''; {\bf 4}- ``Fe~{\sc ii} 4800''; {\bf 5}-
    ``S~{\sc ii} W''; {\bf 6}- ``Si~{\sc ii} 5800''; {\bf 7}-
    ``Si~{\sc ii} 6150''; and {\bf 8}- ``Ca~{\sc ii} IR''.  Short
    vertical lines show the approximate positions where the pseudo-continuum
    is taken in each case. Feature ranges change with time and, due to
    blending, some weaker features are not considered at later
    epochs. Note that after 2-3 weeks past maximum light, the selected
    pseudo-continuum points correspond to emission peaks.  {\it
      Panel}\, ({\bf a}): the region around features \#2 and \#3 for
    near-maximum spectra of \object{SN 1991T} (top), \object{SN 1989B}
    (middle), and \object{SN 1991bg} (bottom).  Feature \#2 is not
    defined in the case of 1991bg-like SNe~Ia because the region is
    dominated by absorption from \ion{Ti}{II}. Adopted feature limits are
    marked with vertical lines. {\it Panel}\, ({\bf b}): an example of
    the pseudo-continuum trace for ``Fe~{\sc ii} 4800'' on a normal SN~Ia near
    the time of maximum light. Here, solid vertical lines show the
    regions where the pseudo-continuum is fitted. Dotted lines mark the
    bounds used to measure the \ew.}
  \label{fig:feat}
\end{figure}

In this analysis, we define an absorption feature as a wavelength
region that is bounded by two local flux maxima. Fig.~\ref{fig:feat}
shows eight such regions, corresponding to the eight strongest
absorption features at optical wavelengths in supernova spectra up to approximately one month after maximum light. Each feature is marked
with a number from 1 to 8 and each number corresponds to a mnemonic
name.  The upper and lower limits of the features vary in time
(because of the SN~Ia envelope expansion), and from SN~Ia to SN~Ia at
a given phase (i.e.~SN~Ia spectral diversity).  We give ranges for these limits in Table~\ref{tab:limits}.

As already mentioned, in order to measure a \ew, a pseudo-continuum must be
determined. We define the pseudo-continuum as the straight line fit
through the two local maxima that bound a feature\footnote{The chosen
maxima are those that maximize the wavelength span of the feature with
the restriction that the derived pseudo-continuum does not intersect the
spectrum within the feature limits with the possible exception of noise
artifacts in case of low signal-to-noise ratio data.} (see panel 
(b) in Fig.~\ref{fig:feat}). A detailed explanation of the
measurement technique is given in  section \ref{ewtec}.

Once the pseudo-continuum is defined, the \ew\ is computed for each
feature within its wavelength limits.  The spectrum is divided by the
pseudo-continuum and the resulting area of the feature is measured (in
units of \AA). In this case, the calculation was approximated by a
simple rectangular integration method:

\begin{equation}
\mbox{\ew}=\sum_{i=1}^N\left(1-\frac{f_\lambda(\lambda_i)}{f_c(\lambda_i)}\right)
\Delta\lambda_i,
\label{eweq}
\end{equation}

\noindent where $\lambda_i$ ($i=1,\ldots, N$) are the central
wavelengths of the bins of size $\Delta\lambda_i$,
$f_\lambda(\lambda_i)$ is the measured flux in bin $i$, and
$f_c(\lambda_i)$ is the fitted pseudo-continuum flux evaluated at the
same points. Deviating points due to bad pixels, night sky residuals,
or narrow host-galaxy lines were rejected using a 3$\sigma$-clipping
algorithm.
 
\begin{table}[htbp]

  \caption{Feature limits.}
  \label{tab:limits}
  \begin{tabular}{llcc}
    \hline
    \hline
    Feature & Mnemonic & Blue-ward & Red-ward \\
    ID & Label & limit range (\AA) & limit range (\AA) \\
    \hline
    1 &  ``Ca~{\sc ii} H\&K'' & 3500 -- 3800 & 3900 -- 4100 \\
    2 &  ``Si~{\sc ii} 4000'' & 3900 -- 4000 & 4000 -- 4150 \\
    3 &  ``Mg~{\sc ii} 4300'' & 3900 -- 4150 & 4450 -- 4700 \\
    4 &  ``Fe~{\sc ii} 4800'' & 4500 -- 4700 & 5050 -- 5550 \\
    5 &  ``S~{\sc ii} W''     & 5150 -- 5300 & 5500 -- 5700 \\
    6 &  ``Si~{\sc ii} 5800'' & 5550 -- 5700 & 5800 -- 6000 \\
    7 &  ``Si~{\sc ii} 6150'' & 5800 -- 6000 & 6200 -- 6600 \\
    8 &  ``Ca~{\sc ii} IR''   & 7500 -- 8000 & 8200 -- 8900 \\
    \hline
  \end{tabular}

\end{table}

The 1$\sigma$ statistical uncertainty was computed by error
propagation from the estimated uncertainties in the spectral flux
($\sigma_{f}$) and in the pseudo-continuum ($\sigma_{c}$):
\begin{equation}
\label{eq:error}
\sigma_{\mbox{\scriptsize \ew}}=\left[\sum_{i=1}^N\left(\frac{\sigma_{f_i}^2(\lambda_i)}
{f_c^2(\lambda_i)}+\frac{f_\lambda^2(\lambda_i)}{f_c^4(\lambda_i)}
\sigma_{c_i}^2(\lambda_i)\right)(\Delta\lambda_i)^2\right]^{1/2}.
\end{equation}

By definition, the rest-frame pseudo-equivalent width is redshift independent
and thus provides a useful tool for comparing spectra from different
objects over a wide range of redshifts.

\subsection{Measurement technique}
\label{ewtec}

Measuring pseudo-equivalent widths on high signal-to-noise ratio
spectra is relatively simple since the local maxima bounding an
absorption feature can be identified easily.  On low signal-to-noise
data the measurements are more difficult. For comparing high redshift
supernovae (which generally have low signal-to-noise) with local SN~Ia
spectra, we have established a measurement technique (to be applied on
spectra regardless of their signal-to-noise ratio) that minimizes
possible systematic effects.

To measure the \ew\ of a spectral feature the local pseudo-continuum must
be determined. To perform this operation we proceeded as follows:

\begin{itemize}
\item The two local maxima that bound the absorption feature are
  visually identified and marked. These are the bounds used in
  performing the sum in Eqn. \ref{eweq}.

\item A small wavelength region (to which
  hereafter we refer to as the fitting region) is selected around each
  identified maximum\footnote{See panel {\bf (b)} in
    Fig.~\ref{fig:feat}, the four vertical lines represent an example
    of the wavelength span of each fitting region.}, always within the
  wavelength ranges listed in Table~\ref{tab:limits}.

\item A straight line is fitted to the data. The result of the fit is taken to be the pseudo-continuum and is
  used in Eqn. \ref{eweq}.
\end{itemize}

The wavelength span of the two fitting regions depends on the
morphology of the local maxima. On the blue end of the spectrum, where
more absorption features are found, the maxima tend to be narrower
than on the red end. Thus, the fitting regions are, on average,
smaller ($\sim$ 20 \AA) in the blue and slightly larger ($\sim$ 25
\AA) in the red part of the spectrum. Larger wavelength spans would
tend to set a lower pseudo-continuum level and thus to produce \ew\ systematically biased toward smaller values.  The size of the
fitting region is independent of the signal-to-noise ratio; however,
in low signal-to-noise ratio spectra, noise spikes that could bias the
pseudo-continuum fit are clipped.

\section{Data sets}
\label{data}
In the following section, we present the high and low redshift data-sets
that we have used in this study.
\subsection{High Redshift Data-set} 
\begin{table*}[th!]
\small
\caption[table data]{A summary of the high-redshift data. (See text
  for details.)}
\begin{center}
\small
\begin{tabular}{lccccccccl}
\hline
 SN-name & Redshift$\tablenotemark{a}$& Date  & Days from$\tablenotemark{b}$& Instrument  & setup & telescope & Exposure& S$/$N$\tablenotemark{c}$&Galaxy$\tablenotemark{d}$\\
  & &(MJD) & B-band  & &  & & time (s) && \%\\
  &  &     & Maximum & &  &  \\
\hline
SN~2001gy &0.511& 52021.3 & $-$7,5(1) & FORS1 & 300V grism + GG435  & VLT-UT1 &  2400 &11&20\\
SN~2002fd $\tablenotemark{e}$&0.279& 52376.1 & $-$7(3) & FORS2 & 300V grism + GG435  & VLT-UT4 &  600&46&28\\
SN~2000fr &0.543& 51676.2 & $-$5.5(1) & FORS1 & 300V grism + GG435  & VLT-UT1 &  7200 &21&16\\
SN~2001gw &0.363& 52021.4 & $-$1(3) & FORS1 & 300V grism + GG435  & VLT-UT1 &  1200&8&19 \\
SN~2001ha &0.58& 52022.0 & 0(3) & FORS1 & 300V grism + GG435  & VLT-UT1 &  3600 &4&6\\
SN~2002gl &0.510& 52413.1 & 0(3) & FORS2 & 300V grism + GG435  & VLT-UT4 &  3000& 9&23\\
SN~2001hc &0.35& 52022.1 & 0(3) & FORS1 & 300V grism + GG435  & VLT-UT1 &  1800 &14&14\\
SN~2001gr &0.540& 52021.0 & 2(3) & FORS1 & 300V grism + GG435  & VLT-UT1 &  3600& 5&57\\
SN~2002gi &0.912& 52407.2 & 2(3) & FORS1 & 300I \  grism + OG590  & VLT-UT3 &  7200&3&37 \\
SN~2001gm &0.478&52021.3 & 5(3)  & FORS1 & 300V grism + GG435  & VLT-UT1 & 2400&3&28\\
SN~2001go &0.552& 52021.3  & 5.6(1) & FORS1 & 300V grism + GG435  & VLT-UT1 &  2400&5&13\\
SN~2002gk &0.212& 52413.3 & 6(3) & FORS2 & 300V grism + GG435  & VLT-UT4 &  900&34&66 \\
SN~2001go &0.552& 52027.1 & 9.5(1) & FORS1 & 300V grism + GG435  & VLT-UT1 &  7200 &10&19\\
SN~2001go &0.552& 52058.1 & 29.5(1) & FORS1 & 300V grism + GG435  & VLT-UT1 &  9000& 4&53\\
\hline
\end{tabular}
\label{tabledata}
\tablenotetext{a}{The redshift is quoted to three significant figures
  if it is determined from host galaxy lines; two significant figures
  when determined from SN~Ia spectral features.}
\tablenotetext{b}{Uncertainties are quoted in parentheses. When the epoch is quoted with 3
  days uncertainty it refers to that of the best matching low-z Ia template.}
\tablenotetext{c}{per 20 \AA\ bin}
\tablenotetext{d}{Estimated fraction of the galaxy light in the SN~Ia spectrum, expressed as percentage of
  the total flux.} 
 \tablenotetext{e}{ 91T-like Type Ia SN}
\end{center}
\end{table*}
\normalsize

\label{sec:hzdata}
The supernova spectra that are analyzed in this work were obtained as
part of several campaigns by the SCP to discover and follow a large
number of SNe~Ia over a wide range of redshifts (see \citet{2005A&A...430..843L} for details). Out of the 20
spectrally confirmed SNe~Ia in \citet{2005A&A...430..843L}, we select
the 12 SNe~Ia ($z$=0.212-0.912) with   signal-to-noise
ratios S/N$>$3 per 20 \AA\ bin to pursue our quantitative
analysis.  One supernova, SN 2001go, was observed at three epochs, so
there are 14 spectra in total.

The biases affecting this sample are complex. First, supernova
searches are magnitude limited, so Malmquist and light curve shape
biases (i.e. brighter/dimmer \sn\ have broader/narrower light curves) make it unlikely that low-luminosity, SN~1991bg-like SNe~Ia
will be found. Also, it is likely that, while ranking supernovae
candidates for follow-up spectroscopy, 1991bg-like SNe~Ia would be
assigned lower priority due to their low luminosity.  Second, it is
more difficult to spectrally confirm high-redshift SNe~Ia in spectra
where host galaxy contamination is above 75\%
\citep{2005A&A...430..843L,2005ApJ...634.1190H}. Although this affects
all SNe~Ia to some degree, lower-luminosity SNe~Ia are more commonly found in
bright ellipticals  Moreover, we can not
exclude that the signal-to-noise cut we performed while choosing the
sub-sample of spectra to study did not introduce an extra selection
bias toward over-luminous objects.  As it will be clear in section
\ref{results}, the spectral indicators we use to search for evolution with
redshift do not unambiguously discern between normal and over-luminous
SN~Ia. The additional scatter that would be introduced by using data
with lower signal-to-noise ratios would obfuscate the result. We make
no attempt to correct the sample for these biases.

\begin{figure*}[ph]
\centering \includegraphics[width=17cm]{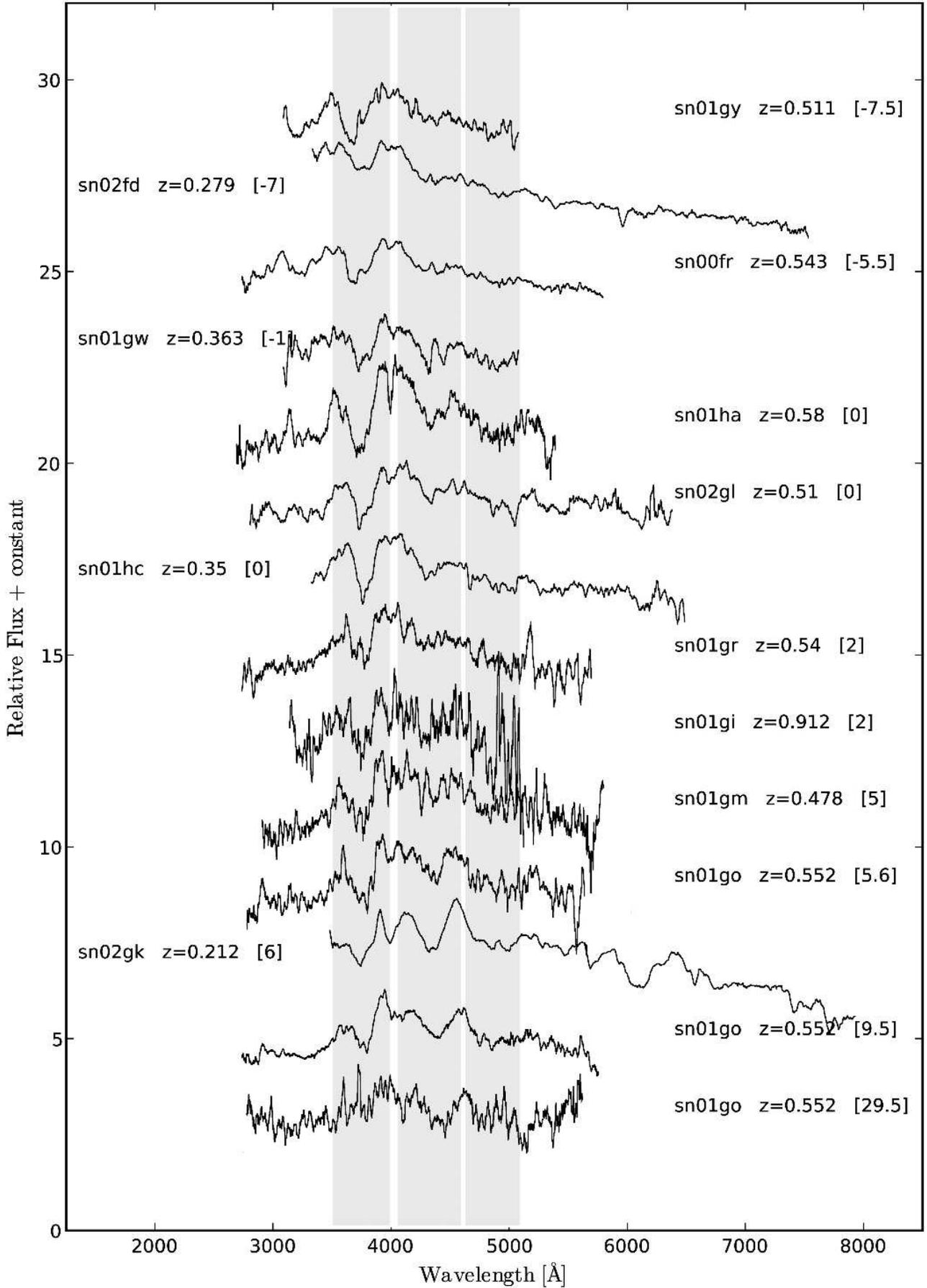}
\caption{ Spectra of the high-redshift SNe~Ia (re-binned to 20 \AA\ )
  used in this study, plotted in rest frame. For each spectrum, we
  indicate the redshift and epoch (in days from B-band maximum; square
  brackets). See Table~\ref{tabledata} for details. Gray vertical
  bands approximately indicate the wavelength regions used for the
  quantitative comparison presented in section \ref{results}.}.
\label{spectra}
\end{figure*}

The spectra, re-binned to 20 \AA\, are shown in Fig.~\ref{spectra},
and are summarized in Table \ref{tabledata}. A full description of the
observations and the data reduction are given in
\citet{2005A&A...430..843L}. For each supernova, we use the error
spectrum, which was estimated from regions free of SN~Ia and host
galaxy light on the sky-subtracted two dimensional spectrum, to
estimate the statistical uncertainties on the quantities we compute in
the following analysis.

High-redshift supernova spectra can contain significant amounts of
light from the host galaxy. On the 2d spectra, the host galaxy and the
SN~Ia are often spatially unresolved, making it difficult to
estimate the contribution of the host to the observed flux.  We
estimated this contribution using a template matching technique based
on a large set of nearby supernovae spectra and galaxy models similar
to those used in \citet{2005A&A...430..843L}.  The contribution of the
galaxy light, relative to the total observed flux, are tabulated in
column 10 of Table~\ref{tabledata}.

The epochs with respect to the $B-$band maximum light  were estimated using the preliminary light
curves, if available, and/or spectroscopic dating by template matching
with low-$z$ SNe~Ia \citep{2005A&A...430..843L,2005AJ....130.2788H}.  The two methods usually agree within three days
\citep{2005AJ....130.2788H,2005ApJ...634.1190H}, therefore we take 3
days to be the uncertainty on the quoted epoch whenever a light curve
estimate of the maximum was not available. The redshift of the
supernova, when quoted with 3 significant figures, was estimated from
host galaxy lines visible in the spectrum. When this was not possible,
the redshift was estimated from supernova spectral features, and is
then quoted with 2 significant figures to account for the large intrinsic
width of SN Ia spectral features.

\subsection{SN~Ia identification}
\label{id}

The identification of  SN~Ia relies primarily on the
detection of the absorption feature at approximately 6150~\AA\ due to
Si~{\sc ii}~$\lambda$6355.  At redshifts above z=0.5, however, this
characteristic feature is redshifted beyond the wavelength range of
most optical spectrographs and the classification of the supernova has
to rely on spectral features that lie at bluer wavelengths
\citep{2005A&A...430..843L,2005AJ....130.2788H,2005AJ....129.2352M}.  Because of the low
signal-to-noise ratio usually available in high-redshift supernova
spectra, this approach is not always conclusive.

We can also use the spectra to identify spectral peculiarities among
SNe~Ia as those found in \taa\ or 91bg-like supernova. In Table
\ref{table_class}, the characteristics of four wavelength regions for
different types and sub-types of supernovae are schematically
reported. Each spectral feature is qualitatively described as {\it
  strong}, {\it weak} or {\it absent} based on the absorption strength
and {\it broad} or {\it narrow} based on the wavelength span.  In the
absence of a procedure that is based on quantitative measurements,
this scheme helps in identifying the SN type and, in the case of
SNe~Ia, the sub-type.

\begin{table*}[htb]
\begin{center}
\small
\caption[tableIaID]{ A description of the spectroscopic features used
  to type SN~Ia at z $\ge$ 0.5.  Four wavelength regions are selected
  for performing the SN~Ia typing. See the text for details.}  \small
\begin{tabular}{lccccc}
\hline Region Id&$\lambda$-Region&Normal Type Ia & 91T/99aa-like& 91bg/86G-like& Type Ib/c \\ 
&Rest Frame[\AA]&&&&\\
\hline 
`Ca {\sc ii} H\&K'$\tablenotemark{a}$&3550-3950&strong/broad&weak or absent/broad$\tablenotemark{b,c}$&strong/broad&evident/broad$\tablenotemark{d}$\\
`Si {\sc ii}'$\tablenotemark{e}$ &3950-4100&evident/narrow$\tablenotemark{c}$&weak$\tablenotemark{b,c}$&absent&absent\\ 

`Fe {\sc ii}'$\tablenotemark{f,g}$ &4600-5200&strong/broad&strong/narrow$\tablenotemark{b}$&strong/broad&strong/broad\\
`S {\sc ii} W'&5200-5600&strong/narrow$\tablenotemark{b,c}$&weak or absent$\tablenotemark{b}$&strong/narrow$\tablenotemark{b,c}$&absent\\ 
\hline
\end{tabular}
\tablenotetext{a}{High and low velocity component in some SNe~Ia.}
\tablenotetext{b}{Around 1 week before maximum light}
\tablenotetext{c}{Around maximum light}
\tablenotetext{d}{A few  exceptions.}
\tablenotetext{e}{Marks the beginning of the distinctive strong Ti {\sc ii} absorption feature in 91bg-like SNe~Ia.}
\tablenotetext{f}{In Normal Ia characteristic line profile time evolution.}
\tablenotetext{g}{Dominated by Fe {\sc iii} in pre-maximum spectra of 91T-like SNe~Ia.}
\label{table_class}
\end{center}
\end{table*}
\normalsize

\subsection{SN~2002fd: A  SN~1991T/SN~1999aa-like Supernova}
\label{sec:02fd}

\begin{figure}[thb]
\centering \includegraphics[width=8cm]{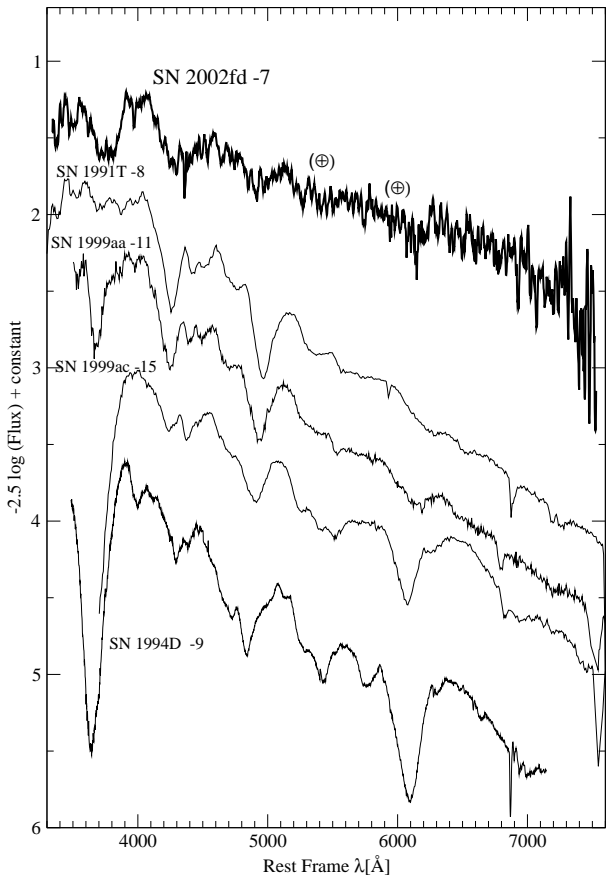}
\caption{SN~2002fd at day $-$7, re-binned to 20 \AA\ per pixel (thick
solid line), compared to normal and peculiar SNe~Ia. The $\oplus$ symbols
mark  regions of strong telluric absorption.}
\label{02fd}
\end{figure}

SN~2002fd ($z=0.279$) is the only supernova in our data set that clearly deviates
from a ``normal Ia" (see Table \ref{table_class}). Within the scheme
described above, the spectrum of SN~2002fd is similar to the spectra
of SN~1999aa \citep{2004AJ....128..387G}, a \taa\ supernovae
(Fig.\ref{02fd}).  The strength of the `Ca~{\sc ii}~H\&K' feature in
SN~2002fd is stronger than in \taa\ \sne\ and weaker than in normal
\sne.  The `Fe~{\sc ii}' and `S~{\sc ii} W' regions are similar to
those in SN~1999ac \citep{2005AJ....130.2278G}. Given the
low-redshift, Si~{\sc ii}~$\lambda$~6355 is also visible.  This
feature in SN~2002fd is intermediate in strength compared with
SN~1999aa and in normal SNe~Ia. From this qualitative analysis of the
spectrum, we classify SN~2002fd as a peculiar \sn, similar to the
\taa\ SN~1999aa.  In section \ref{02fdhk}, we show that the
pseudo-equivalent widths of the absorption features in SN~2002fd are also
consistent with those found at low redshift for the \taa\ objects.

Finding SNe~Ia with spectral characteristics similar to those of
SN~1991T at high redshifts is important. The lack of such SNe in high
redshift surveys might be a sign of evolution.  In a distance limited
survey, \citet{2001ApJ...546..734L} found that approximately 20\% of
the analyzed data set of nearby SNe~Ia could be classified as
SN~1991T/SN~1999aa-like which were peculiar, over-luminous SNe~Ia
\citep{2004AJ....128..387G}.  
This percentage is probably higher than that
of peculiar \taa\ found at high-redshift.  Currently, the reported fraction of 91T-like SNe Ia is $\leq 5 \%$  \citep{2005A&A...430..843L,2005AJ....129.2352M}.
 However, the fraction of these
SNe~Ia in high-redshift surveys is uncertain because of  the difficulty in
identifying such SNe~Ia.   Several more \taa\ SNe~Ia may have already been
observed at high-redshift but not clearly identified because  the spectrum was
taken well after maximum light or the light-curve had a \dmft\  consistent with normal \sne.
Over-luminous \sne\ such as \taa\ SNe~Ia generally
have broader light-curves than normal \sne. However, it is now
becoming evident that \taa\ SNe~Ia do not always have high \dmft\ values and that broad light-curve \sne\ do not always show the
spectra peculiarities seen in \taa\ SNe~Ia. Example of the latter case
are SN~1999ee \citep{2002AJ....124..417H}, SN~2002cx \citep
{2003PASP..115..453L} or SN~1999aw \citep{2002AJ....124.2905S}.
\citet{2001ApJ...546..719L} computed that the peculiarity rate for
\taa\ SNe~Ia should vary between 6\% and 18.6\%.  In this case the
observed peculiarity rate at high redshift (about 4\% in
\citet{2005AJ....129.2352M} and  about 5\% in our data set) is 
consistent with the computations, especially if one considers that the
expected number of peculiar supernovae is highly dependent on the
 search strategy.  Nevertheless, the identification of
peculiar SNe~Ia in high-redshift samples is an important step toward
determining whether the range of SNe~Ia sub-types that is observed at
low redshift is also observed at high redshifts.
Of course,  large data samples with high signal-to-noise ratio spectra are required to check whether the
fraction of peculiar high-redshift SNe~Ia is consistent with that
found in the low-redshift Universe or if there is an evolution with
redshift in the relative fractions, which might affect the derivation
of cosmological parameters from SNe~Ia.

\subsection{Nearby Supernova Data-sets}
\label{lowzdata}
Two different data sets are used to derive the properties of \ew\ in
local SNe~Ia.

{\bf Set~A} - 77 spectra (presented in \citet{gastonthesis}) from 13
of the SNe~Ia discovered and followed by the Supernova Cosmology
Project ({\sc scp}) in collaboration with members of the {\sc eros}
\citep{2000A&A...362..419H}, {\sc quest} \citep{1999ApJ...524L.103S},
and Nearby Galaxies SN Search \citep{1999IAUC.7130....1G} teams. The
redshift range of these SNe~Ia is $0.01<z<0.15$.

\noindent The two-dimensional raw images were reduced according to
standard procedures. The calibrated spectra were additionally
corrected for atmospheric and Galactic extinction
\citep{1989ApJ...345..245C,1998ApJ...500..525S} and their
flux-calibration was checked against measured broad-band photometry
and found in agreement within the quoted uncertainties. The spectra
were de-redshifted.  More details about these data can
be found in Table~\ref{tab:low-z}. All the spectra in {\bf Set~A}
include estimated statistical uncertainties for each wavelength bin.
Host-galaxy light was present in a subset of these spectra. 
This contribution was
estimated and subtracted (for details on the procedure see
Sec. \ref{sec:hzdata}) in those cases where it exceeded 10\% of the
total flux.

\begin{table*}[htbp]
  \begin{center}
    \caption{SNe~Ia from {\bf Set~A} ({\sc scp} data: \citet{gastonthesis}).}
    \label{tab:low-z} 
    \begin{tabular}{lcccl}
      \hline
      \hline
      & & Host & Galaxy & Rest-Frame \\
      SN & $z$ & Galaxy & Type$^{\mathrm{a}}$ &
      Spectral Epochs$^{\mathrm{b}}$ \\ 
      \hline
      1999aa & 0.0144 & NGC 2595 & SAB(rs)c & $-$11, $-$7, $-$3, $-$1, 5, 6, 14,
      19, 25, 28, 33, 40, 47, 51, 58 \\
      1999ac & 0.0095& NGC 6063 & Scd & $-$15, $-$9, 0, 2, 8, 11, 16, 24,
      28, 31, 33, 39, 42 \\
      1999ao & 0.054 & Anon. & S: & 5, 7, 10, 13, 18, 34, 40 \\
      1999au & 0.124 & Anon. & S: & 12, 17 \\
      1999av & 0.05 & Anon. & E/S0: & 2, 5, 9, 31 \\
      1999aw & 0.038 & Anon. & (?) & 3, 5, 9, 12, 16, 24, 31, 38 \\
      1999be & 0.019 & Anon. & (?) & 14, 19, 26, 33, 37, 44 \\
      1999bk & 0.096 & Anon. & E/S0: & 4, 7, 9 \\
      1999bm & 0.143 & Anon. & S: & 3, 6, 21 \\
      1999bn & 0.129 & Anon. & S: & 2, 14, 22 \\
      1999bp & 0.077 & Anon. & S: & $-$2, 0, 1, 6, 17, 23 \\
      1999bq & 0.149 & Anon. & E/S0: & 3, 18 \\
      1999by & 0.0021 & NGC 2841 & SA(r)b & 1, 6, 16, 27, 34 \\
      \hline
    \end{tabular}
  \begin{list}{}{}
  \item[$^{\mathrm{a}}$] Hubble type of the host galaxy. An entry
  followed by a colon is a classification based on the host galaxy
  spectrum. The rest are taken from NED\footnotemark 
  \item[$^{\mathrm{b}}$] Rest-frame days since $B$-band maximum light.
  \end{list}
  \end{center}
\end{table*}
\begin{table*}[htbp]
  \begin{center}
  \caption{SNe~Ia from {\bf Set~B} (Public data).}
  \label{tab:public}
  \begin{tabular}{lcccll}
    \hline
    \hline
    & $z$ & Host & Galaxy  &  Rest-Frame& \\
    SN &  & Galaxy & Type$^{\mathrm{a}}$ &
    Spectral Epochs$^{\mathrm{b}}$ & References\\ 
    \hline
    1981B & 0.0060 & NGC 4536 & SAB(rs)bc & 0, 17, 26, 29, 35, 49 & 1 \\
    1986G & 0.0018 & NGC 5128 & S0 pec & $-$7, $-$5, $-$1, 0 ,3, 21,
    28, 41, 44, 55 & 2\\
    1989B & 0.0024 & NGC 3627 & SAB(s)b &
    $-$7, $-$5, $-$3, $-$2, $-$1, 3, 5, 8, 9, 11, 16, 18, 19 & 3 \\
    1990N & 0.0034 & NGC 4639 & SAB(rs)bc & $-$14, $-$7, 7, 14,
    17, 38 & 4, 5, 6 \\
    1991bg & 0.0035 & NGC 4374 & E1 & $-$2, 0, 13, 16, 
    23, 24, 30, 31, 44, 52 & 7, 8 \\
    1991T & 0.0058 & NGC 4527 & SAB(s)bc &
    $-$11, $-$9, $-$8, $-$7, $-$6, $-$5, $-$3, 0, 10, 15, 23, 24, 42,
    45 &  9, 10, 11, 12 \\ 
    1992A & 0.0063 & NGC 1380 & SA0 &
    $-$5, $-$1, 3, 7, 9, 11, 16, 17, 24, 28, 37, 46 & 13 \\
    1994D & 0.0015 & NGC 4526 & SAB(s) &
    $-$10,$-$9,$-$8,$-$7,$-$5,$-$3, 2, 4, 6, 8, 11, 13, 14, 16,
    18, 20, 25 & 14, 15 \\ 
    \hline
  \end{tabular}
  \begin{list}{}{}
  \item[$^{\mathrm{a}}$] Hubble type of the host galaxy from
  NED\footnotemark[\value{footnote}].
  \item[$^{\mathrm{b}}$] Rest-frame days since $B$-band maximum light.
  \item[Sources:] (1)~\citet{1983ApJ...270..123B}; (2)~\citet{1987PASP...99..592P};
  (3)~\citet{1994AJ....108.2233W}; (4)~\citet{1991ApJ...371L..23L}; (5)~\citet{1992AJ....103.1632P};
  (6)~\citet{1993A&A...269..423M}; (7)~\citet{1992AJ....104.1543F}; (8)~\citet{1993AJ....105..301L};
  (9)~\citet{1992ApJ...384L..15F}; (10)~\citet{1992ApJ...387L..33R};
  (11)~\citet{1992AJ....103.1632P}; (12)~\citet{1992ApJ...397..304J}; 
  (13)~\citet{1993ApJ...415..589K}; (14)~\citet{1996MNRAS.281..263M}; (15)~\citet{1996MNRAS.278..111P}.
  \end{list}
  \end{center}
\end{table*}

{\bf Set~B} - 89 published spectra from 8 well-observed, nearby
objects (see Table~\ref{tab:public}). Additionally, the spectra were scaled to match the
photometry. Given the absence of
published uncertainties on these spectra, statistical errors were
estimated from the pixel-to-pixel variation.\\

The epochs used in this analysis are based on light curve
estimates. In the case of {\bf Set~A}, the date of maximum $B$-band
brightness was determined using preliminary light curves. Therefore
the epochs used to date the spectra were taken as the integer number
of days since maximum light.  The photometric data available for both
sets were used to fit the template $B$-band light curve given by
\citet{2001ApJ...558..359G} and thus to obtain the values of the
light curve decline parameters $\Delta m_{15}(B)$ and stretch ($s$) for each
SN~Ia.

The two sets contain some peculiar SNe~Ia, including the prototypes of
the two subclasses: \object{SN 1991T} and \object{SN 1991bg}.
\object{SN 1999aa} \citep{2004AJ....128..387G}, \object{SN 1999aw}
\citep{2002AJ....124.2905S}, and \object{SN 1999bp}
\citep{gastonthesis} are included in the 1991T-like subclass. All
these SNe~Ia present values of the decline rate parameter $\Delta
m_{15}(B)<1.0$ ($s>1.1$) and thus have a slow post-maximum decline in
luminosity. At the other extreme, \object{SN 1986G}
\citep{1987PASP...99..592P,1992A&A...259...63C} and \object{SN 1999by}
\citep{2001AJ....121.3127V,2004ApJ...613.1120G} belong to the
1991bg-like subclass. These are fast-declining \sne, with $\Delta
m_{15}(B)>1.70$ ($s<0.80$).  The case of \object{SN 1999ac}
\citep{2005AJ....130.2278G} is considered separately. This SN~Ia has
photometric and spectroscopic peculiarities that make it a unique
object: its light curve shows a slow rise similar to \object{SN 1991T}
but a fast decline \citep{Phillips:2002cg} and its spectrum is similar
to \object{SN 1999aa}.

\footnotetext{The NASA/IPAC Extragalactic Database (NED) is operated
  by the Jet Propulsion Laboratory, California Institute of
  Technology, under contract with the National Aeronautics and Space
  Administration.}

\section{Results}
\label{results}
\noindent We first perform a qualitative analysis (sections
\ref{vel} and \ref{sec:local}) and then turn to a more quantitative approach
(sections \ref{02fdhk} and  \ref{stat}).  

\subsection{Absorption Velocities}
\label{vel}

\begin{figure}
\centering \includegraphics[width=8cm]{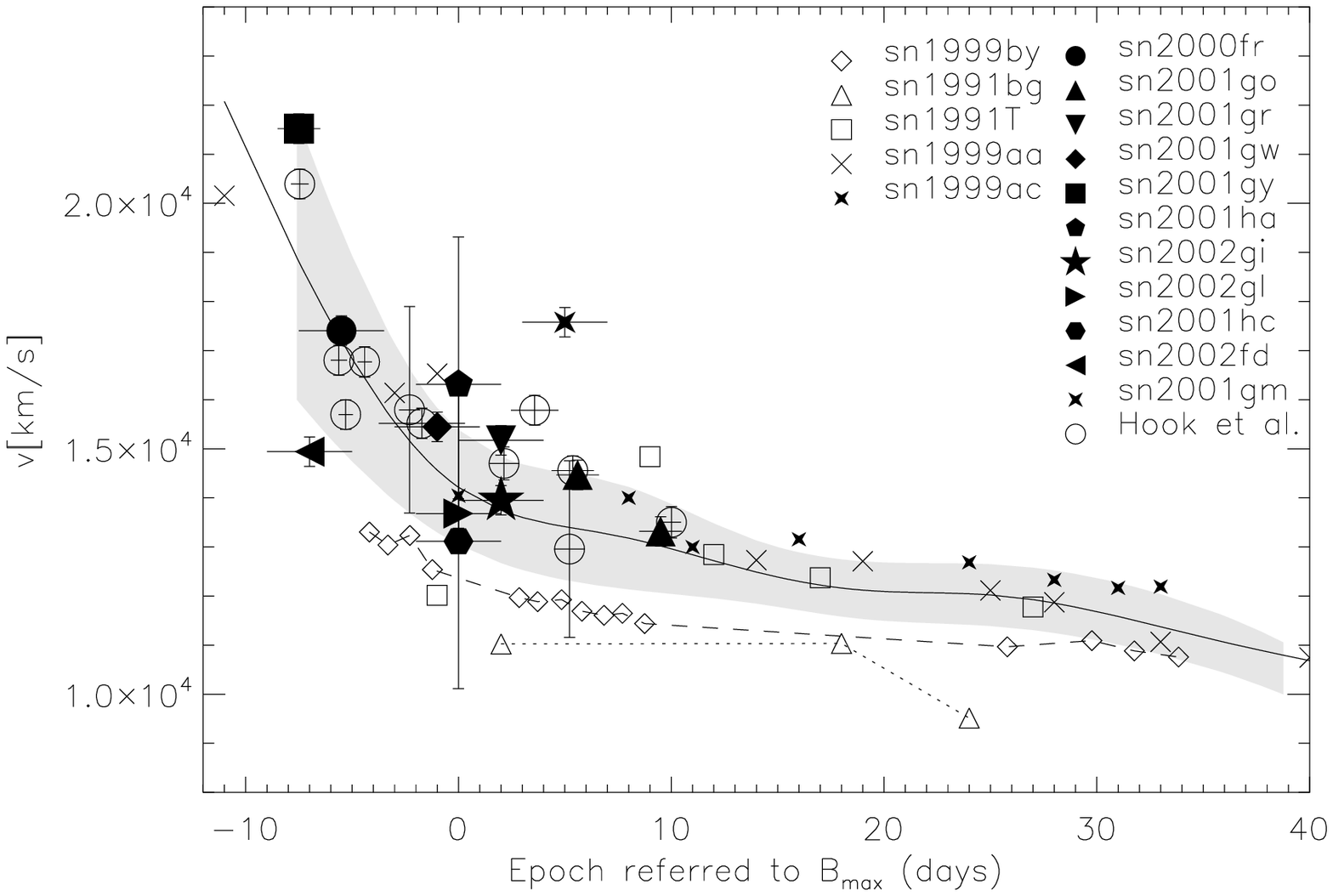}
  \caption{ The change in the blueshifts of the Ca~{\sc ii}~H\&K absorption minimum 
    with epoch for a sample of high-redshift SNe~Ia (presented in section
    \ref{data}, filled symbols, and from \citet{2005AJ....130.2788H} open circles) and a
    sample of low-redshift SNe~Ia. The dashed and dotted lines indicate
    the values of extremely under-luminous SNe~Ia SN~1999by
    \citep{2004ApJ...613.1120G} and SN~1991bg
    \citep{1993AJ....105..301L} respectively. The solid line indicates
    the average trend for Ca~{\sc ii}~H\&K, which has been computed
    using a large data set of low-redshift normal SNe~Ia
    \citep{2004AJ....128..387G}. The gray band shows the dispersion (1
    standard deviation) of the data about the average trend.}
  \label{ca_vel}
\end{figure}

For normal SNe~Ia, the magnitude of the blueshift of the Ca~{\sc ii}~H\&K 
absorption minimum  drops rapidly,
from values around 22000 km/s before maximum light to 14000 km/s at
maximum light. After maximum light, the decline in velocity flattens
and decreases by about 4000 km/s in 50 days. The mean trend for normal
SNe~Ia, from 10 days before maximum light to 40 days after maximum
light, is shown in Fig.~\ref{ca_vel} together with the evolution in
the velocities of fast and slow-declining local SNe~Ia.  The shaded area represents the 1~$\\sigma$ dispersion about
the mean for normal SNe~Ia.
The trend and the dispersion have been computed
from a large sample of nearby supernova \citep{2004AJ....128..387G}.

The measured absorption velocity of the ejecta in 
 under-luminous SNe~Ia (i.e.~SN~1999by and SN~1991bg plotted with the
dotted and dashed lines, respectively) is approximately 1.5 sigma lower
than the mean absorption velocity in normal SNe~Ia. Therefore, the
measurement of a low blueshift of the Ca~{\sc ii}~H\&K absorption minimum cannot be
used to identify a SNe~Ia as under-luminous as already pointed out in
\citet{2006AJ....131.1648B}. Peculiar over-luminous SNe~Ia such as
SN~1991T and SN~1999aa show blueshifts of Ca~{\sc ii}~H\&K absorption minimum that are
within one sigma of those of normal SNe~Ia. SN~1991T at one day before
maximum light shows a blueshifts comparable with that of SN~1999by, a fast-declining SN
Ia.

In Fig.~\ref{ca_vel}, the velocities of both the low-redshift and
high-redshift SNe~Ia are individually measured by performing an
error-weighted non-linear least-squares fit to the entire line profile,
manually selecting the end points limiting the wavelength region where
to perform the fit. The line profile is modeled with a Gaussian plus a
linear component. This method accurately reproduces the absorption line
profile and it has been successfully used in previous studies
(e.g. \citet{2005AJ....130.2278G,2005AJ....130.2788H}). All the SNe~Ia
in our data set were measured (Table \ref{veldata}), with the exception of SN~2001gk, for
which the line profile is incomplete, and SN~2001go at +29 days, for
which the signal-to-noise is too low to correctly identify the
absorption. The uncertainty in the redshifts is taken to be
$ \sigma_{cz}=300\ km/s$ if the redshift was
estimated from galaxy lines.  For SN~2001ha and SN~2001hc, we could
not identify host galaxy lines, so their redshifts were estimated by
comparing their spectra with the spectra of nearby SNe~Ia \citep{2005A&A...430..843L}. In these
cases the uncertainty is increased to
$\sigma_{z}=0.01\equiv\sigma_{cz}=3000\ km/s$.  This uncertainty
dominates the statistical uncertainty from the fit.  We note that the
velocities of our high-redshift SNe~Ia are consistent with the main
trend of spectroscopically normal local SNe~Ia. In Fig~\ref{ca_vel}
open circles indicate the measurements reported in
\citet{2005AJ....130.2788H} where a similar result was found in an
independent sample of high redshift SNe~Ia. 

\begin{table*}[htb]
\begin{center}
\small
\caption{Magnitude of the blueshift  of Ca~{\sc ii}~H\&K absorption minimum.  Measurement uncertainties are reported in parenthesis.}
\begin{tabular}{llr}
\hline
SN &day&vel [\kmps] \\
\hline
sn2001gy& $-$7.5 (1) &21520 (300)    \\
sn2002fd& $-$7 (2)& 14940 (300) \\
sn2000fr& $-$5.5 (1) &17400 (300) \\
sn2001gw& $-$1 (2) &15440 (300)     \\
sn2002gl& 0 (2)& 13680 (300)\\
sn2001ha& 0 (2) &16310 (3000) \\
sn2001hc& 0 (2) &13110 (3000) \\ 
sn2002gi& 2 (2) &13950 (300)     \\
sn2001gr& 2 (2)& 15170 (300)\\
sn2001gm& 5 (2)& 17580 (300)\\
sn2001go& 5.6 (1)& 14470 (300)\\
sn2001go& 9.5 (1) &13320 (300)\\
\hline 
\end{tabular}
\label{veldata}
\end{center}
\end{table*}

\subsection{Pseudo-Equivalent Widths}
\label{sec:local}
Given the mean redshift of the
high-redshift sample (z=0.49) and the typical S/N ratio of the
spectra, we restrict our analysis to the bluest and strongest features
- \#4 ``Fe~{\sc ii} 4800'', \#3 ``Mg~{\sc ii} 4300'' and \#1 ``Ca~{\sc
  ii} H\&K''.

\subsubsection{Estimating Systematic Errors}
\label{subsec:syst}
\label{ewsys}

Possible systematic errors arising from the choice of the fitting
region at either side of the feature where the pseudo-continuum is
fitted were accounted for by randomly shifting these regions
(typically within a third of the region size in each
direction\footnote{Considering the uncertainty on the location of a
  region of size $l$ to be $l/\sqrt{12}\sim l/3$, as derived from the
  second moment of a uniform distribution.})  and computing the
weighted root-mean square deviation (rms) of the measured \ew s.
This was the dominant source of uncertainties when the signal-to-noise
ratio per resolution element was above $\sim$10. Additionally, for
lower signal-to-noise spectra (specifically in the case of our high
redshift data-set) the central wavelength of the fitting region was
randomly shifted according to a Gaussian distribution with
$\sigma$=10\AA. The change in the \ew\ is insignificant for high
signal-to-noise ratio data.  The standard deviations
of these distributions were
chosen in a conservative manner so as to take into account even large
systematic effects.  This source of error was added quadratically to
the one given in Eqn.~\ref{eq:error}.

It is known that \ew s can be affected by poor resolution and low
signal-to-noise ratios (see, for example
\citet{1992Sci...257.1978G}). Both effects were tested. Boxcar
smoothing was used to decrease the resolution of the best-sampled
spectra ($\sim$10
\AA/pixel) so that the range of resolutions in the present data set were
tested. Due to the large intrinsic width of the broad SN~Ia features,  no significant change in the measured \ew\ was found.

Signal-to-noise ratios ranged from about 5 to several
hundred. When Gaussian noise was added to the best-quality spectra, in
order to reproduce that quality range, no significant bias was
detected in the resulting \ew s.
 
Additionally, the effect of reddening was tested by artificially
adding up to $E_{B-V}=0.32$ magnitudes of reddening (corresponding to
$A_V=1$ mag, with $R_V=3.1$) following the law given in
\citet{1989ApJ...345..245C}. This produced no significant change in
the resulting \ew s.  This is expected
since the \ew s are defined over a limited wavelength range.

Further systematic effects could arise from host-galaxy light that has
not been perfectly removed. The effect of additional signal underlying
the SN~Ia spectrum would be to lower the \ew. The spectra
from {\bf Set B} in the present sample correspond to very bright,
nearby SNe~Ia, for which SN~Ia and host-galaxy spectra can be
resolved. For more distant SNe~Ia, from {\bf Set A} and the high
redshift data set, the light from the host can contribute up to 50\%
of the light in the extracted spectra. We have tested how errors
in estimating the amount of host galaxy light can affect the
\ew s of a typical near-maximum light SNe~Ia spectrum. Template
galaxy spectra of Hubble types E and Sc were added to the SN~Ia
spectrum in order to simulate contamination levels ranging up to
50\% of the total integrated flux between $3500$ and $9000$ \AA. The
\ew s of the features were then measured on every spectrum. The
relative decrease in the \ew\ with increasing contamination levels was found to
be approximately linear. Table~\ref{tab:hcont} summarizes these
results by giving the relative decrease in the \ew\ per each 10\% of
unaccounted contamination on the total integrated flux for both
galaxy types.  Since SNe~Ia near maximum light are generally bluer
than their hosts, errors in estimating the amount of host galaxy
contamination leads to larger errors in the \ew s for
features at redder wavelengths.  For early type galaxies, because of
the presence of the Balmer break around 4000\AA, the effect on the
\ew\ of ``Ca~{\sc ii} H\&K'' is less than 10\% even for 50\%
contamination.

\begin{table}[htbp]
  \begin{center}
  \caption{The fractional decrease in the \ew\ corresponding to a 10\% increase in the
amount of contamination from the host.}
  \label{tab:hcont}
  \begin{tabular}{llcc}
    \hline
    \hline
    Feature & Mnemonic & \multicolumn{2}{c}{Host Type} \\
    ID      & Label     & E$^{\mathrm{a}}$ & Sc$^{\mathrm{a}}$ \\
    \hline
    1 &  ``Ca~{\sc ii} H\&K'' & 0.019 & 0.080 \\
    2 &  ``Si~{\sc ii} 4000'' & 0.048 & 0.074 \\
    3 &  ``Mg~{\sc ii} 4300'' & 0.037 & 0.073 \\
    4 &  ``Fe~{\sc ii} 4800'' & 0.070 & 0.074 \\
    5 &  ``S~{\sc ii} W''     & 0.112 & 0.074 \\
    6 &  ``Si~{\sc ii} 5800'' & 0.128 & 0.066 \\
    7 &  ``Si~{\sc ii} 6150'' & 0.103 & 0.084 \\
    8 &  ``Ca~{\sc ii} IR''   & 0.155 & 0.114 \\    
    \hline
  \end{tabular}
   \begin{list}{}{}
  \item[$^{\mathrm{a}}$]  Fractional decrease in \ew.
  \end{list}
  \end{center}
\end{table}

\subsubsection{``Fe~{\sc ii} 4800'' (\#4).}
\label{Fe}

\begin{figure}[htbp]
  \centering
  \includegraphics[width=\hsize]{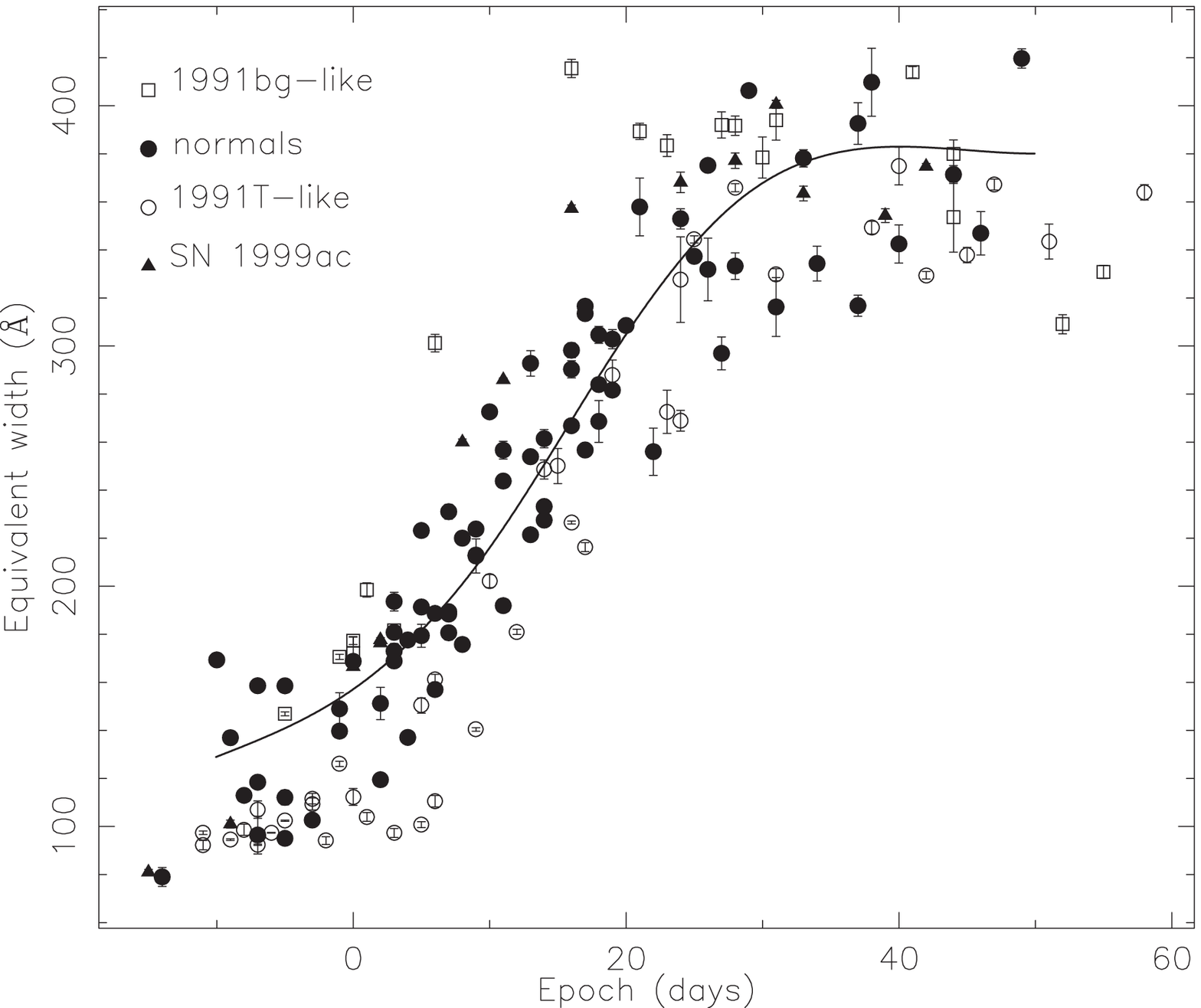}
\centering \includegraphics[width=\hsize,bb= 80 36 539 370]{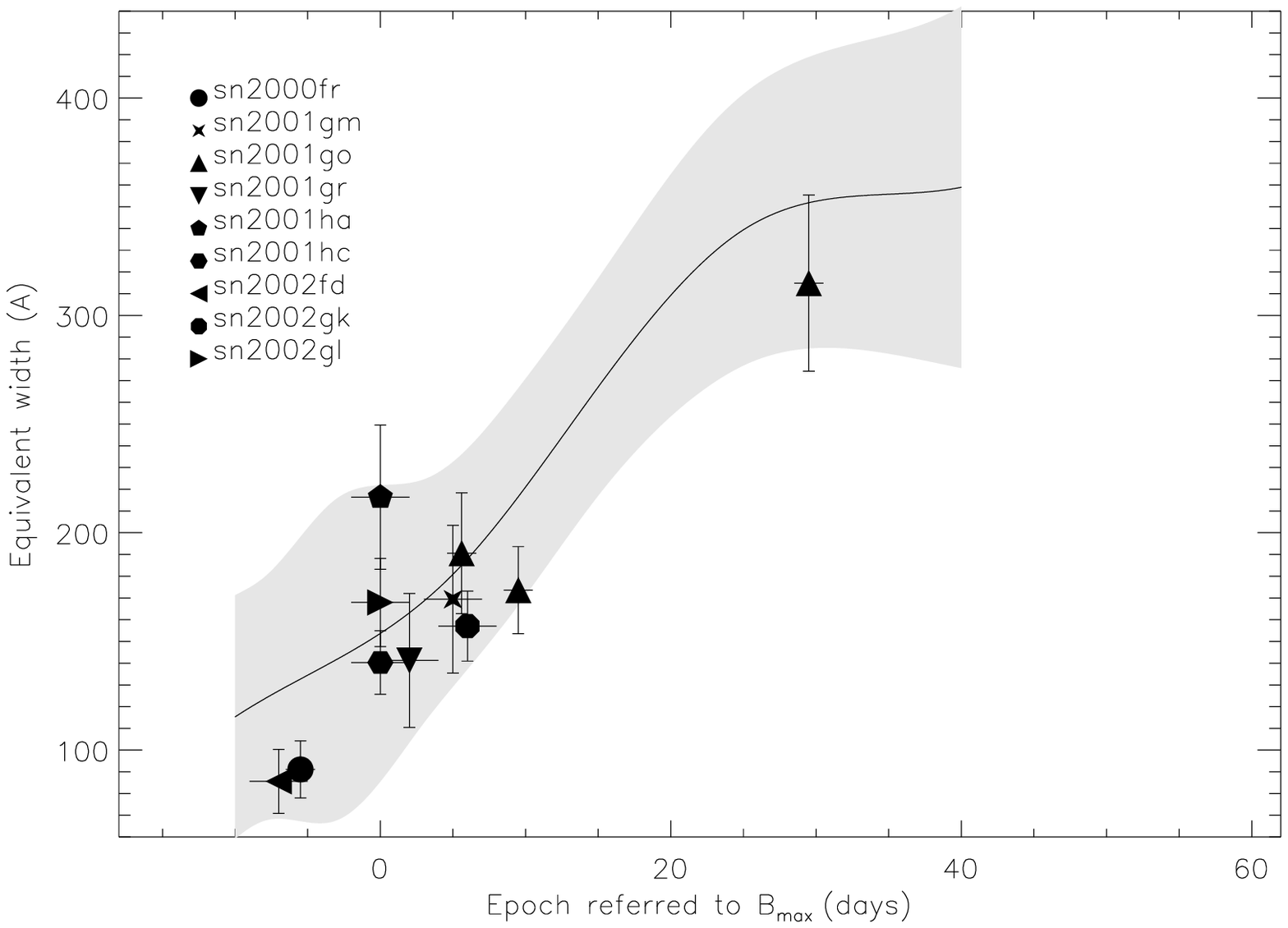}

  \caption{\ \ \ \ \ \ \ \ \ \ \ \ \ \ \ \  \ \ \ \ \ \ \ \ \ \ {\bf``Fe~{\sc ii} 4800''} \newline
  {\it Upper Panel: }Measured pseudo-equivalent width
    corresponding to the ``Fe~{\sc ii} 4800'' feature (\#4). SN
    1991bg-like objects are marked with open squares, 1991T-like
    SNe~Ia with open circles, normal SNe~Ia with filled circles and
    \object{SN 1999ac} with triangles.  Error-bars include the error
    described by Eqn.~\ref{eq:error} as well as systematic
    uncertainties arising from the pseudo-continuum fit. The solid
    line shows a cubic spline function used to represent the average
    evolution of normal SNe~Ia between days $-10$ and
    $+50$. In general, 1991bg-like SNe~Ia lie above the average curve
    whereas 1991T-like SNe~Ia lie below it.  {\it Lower Panel: }A
    comparison between the `Fe~{\sc ii}~4800' \ew s in low- and
    high-redshift SNe~Ia.  High-redshift supernova are indicated by
    large filled symbols.  The gray filled area represents the 95\%
    probability region for normal low-redshift SNe~Ia. See text for details.}
  \label{fig:Femod}
  \label{fig:FeII}
\end{figure}

The evolution in the \ew\ of feature \#4, ``Fe~{\sc ii} 4800'', in
nearby supernovae is shown in Fig.~\ref{fig:Femod}.  The number of
data points enable us to compute a mean trend, which is shown as the
solid line in Fig.~\ref{fig:Femod},  for normal \sne\ only.  The
curve was built in the range $-10\mbox{d} < \mbox{epoch} < 50\mbox{d}$
by dividing the epochs into 5 bins, calculating a weighted average of
the \ew\ in each bin, and finally tracing the spline function through
those points as a general indication of the followed time evolution. The coordinates (epoch,\ew), in units of days and \AA,
of the 5 points defining the curve are: (-5,134); (5,181); (15,267);
(25,339); (35,356).

The \ew\ of ``Fe~{\sc ii} 4800'' monotonically increases with phase
from before maximum light to about 30 days after maximum light. 
This is due to the increasing optical
depth of Fe~{\sc ii} lines from around maximum light onward
and to the subsequent overlapping of several Fe~{\sc ii} lines from
around 15 days after maximum light (see Fig.~\ref{fig:feat}).

The evolution of \ew s is similar for all
Ia subtypes, but offset from the
mean curve. In general, 1991bg-like SNe~Ia lie above the curve and
1991T-like SNe~Ia lie below it but there is no firm correlation with \dmft.  This is summarized in
Table~\ref{tab:FeII_disp} which shows the distribution of the three
SNe~Ia subtypes with respect to the average curve. 

We note that for normal type~Ia SNe the average trend in the first
data bin is based on four data points from SN~1994D and one each from
SN~1989B and SN~1990N. The mean is then biased toward the average
\ew\ value of SN~1994D between -10 and -6 days.  This is higher than
the measurement obtained on the other SNe~Ia, thus, the pre-maximum
average trend is biased toward the trend observed in SN~1994D. 
More
\ew\ measurements are needed to firmly establish the pre-maximum
evolution in normal SNe~Ia.

\begin{table*}[htb]
  \begin{center}
  \caption{Dispersion of ``Fe~{\sc ii} 4800'' \ew\ for the three
  SNe~Ia subtypes. See text for details.}
  \label{tab:FeII_disp}
  \begin{tabular}{l@{\hspace{2.7em}}ccc@{\hspace{2.7em}}cc@{\hspace{2.7em}}cc}
    \hline
    \hline
    Epoch & \multicolumn{3}{c@{\hspace{2.7em}}}{Normal} &
    \multicolumn{2}{c@{\hspace{2.7em}}}{1991T-like} &
    \multicolumn{2}{c@{\hspace{2.7em}}}{1991bg-like} \\ 
    ~~bin & $n$ & $\langle\mbox{\ew}\rangle$ & 
    $\delta$ & $n$ & 
    $\delta_{\mbox{\scriptsize \ew}}$ & $n$ & 
    $\delta_{\mbox{\scriptsize \ew}}$ \\ 
     & & (\AA) & (\AA) &  & (\AA) &  & (\AA)\\
    $(1)$ & $(2)$ & $(3)$ & $(4)$ & $(5)$ & $(6)$ & $(7)$ & $(8)$\\
    \hline
    $[-10,-6]$ & 6  & $132\pm11$ & 28 & 5 & $-29\pm3$  & 0 & $\cdots$ \\
    $[-5,-1]$  & 6  & $114\pm12$ & 37 & 5 & $-36\pm5$  & 2 & $16\pm4$ \\
    $[0,5]$    & 12 & $164\pm10$ & 28  & 5 & $-62\pm9$  & 4 & $24\pm7$ \\
    $[6,11]$   & 15 & $200\pm8$  & 25  & 4 & $-58\pm15$ & 1 &
    $113$$^{\mathrm{a}}$ \\
    $[12,18]$  & 15 & $265\pm9$  & 25  & 5 & $-47\pm12$ & 1 &
    $139$$^{\mathrm{a}}$ \\
    $[19,27]$  & 10 & $316\pm12$ & 30 & 5 & $-34\pm18$ & 3 & $58\pm8$ \\
    $[28,40]$  & 9  & $353\pm13$ & 36 & 5 &   $3\pm9$  & 3 & $37\pm5$ \\
    \hline
  \end{tabular}
  \end{center}
  \begin{list}{}{}
  \item[]
  Columns: (1) Range of epoch bins in days; (2) Number of points from
   normal SNe~Ia; (3) Average \ew\ for  normal SNe~Ia; (4)
  Dispersion (rms) of normal SNe~Ia around the average curve\footnotemark; 
  (5) Number of points from 1991T-like SNe~Ia; (6) Mean deviation of
  1991T-like SNe~Ia from the average curve. Uncertainties do not include the
  computed dispersion of normal SNe~Ia around the curve ($\delta$);
  (7) Number of points from 
  1991bg-like SNe~Ia; (8) Mean deviation of 1991bg-like SNe~Ia from the average
  curve. Uncertainties do not include the
  computed dispersion of normal SNe~Ia around the curve ($\delta$).
  \item[$^{\mathrm{a}}$] Only one measurement, thus no uncertainty is
    given.
  \end{list}
\end{table*}

In Fig~\ref{fig:FeII} (lower panel) the ``Fe~{\sc ii}~4800'' \ew s of
the high redshift sample are shown.  All SNe~Ia (with the exception of
SN~2001gu, SN~2001gw, SN~2001gy and SN~2002gi, for which the
absorption feature was not easily identifiable) were found to lie
within the 95\% probability distribution of low-redshift supernovae
indicated by the gray filled area.

\subsubsection{``Mg~{\sc ii} 4300'' (\#3)}
\label{MgII}

\begin{figure}[htbp]
  \includegraphics[width=\hsize]{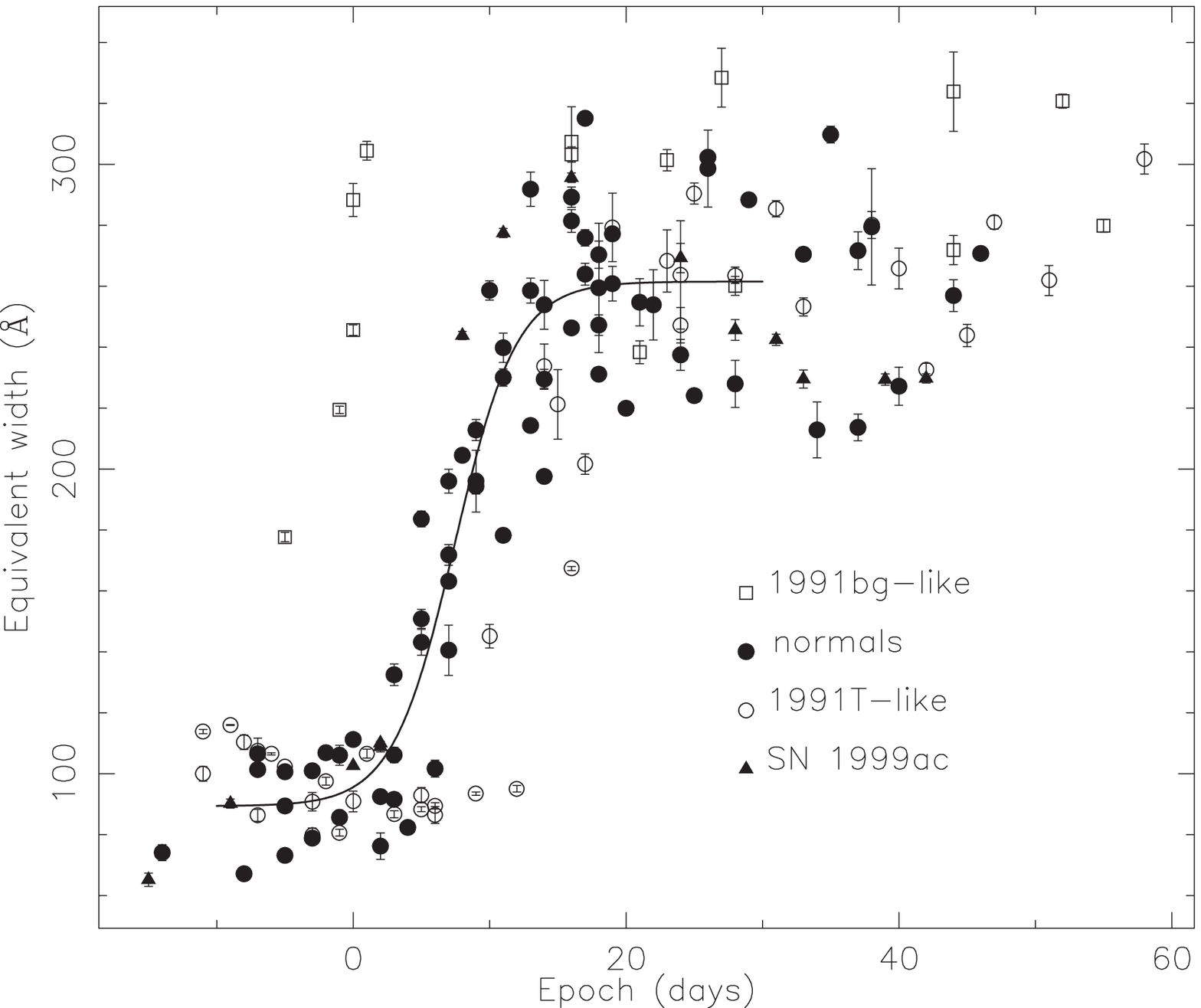}
\centering \includegraphics[width=\hsize,bb= 80 36 539 370]{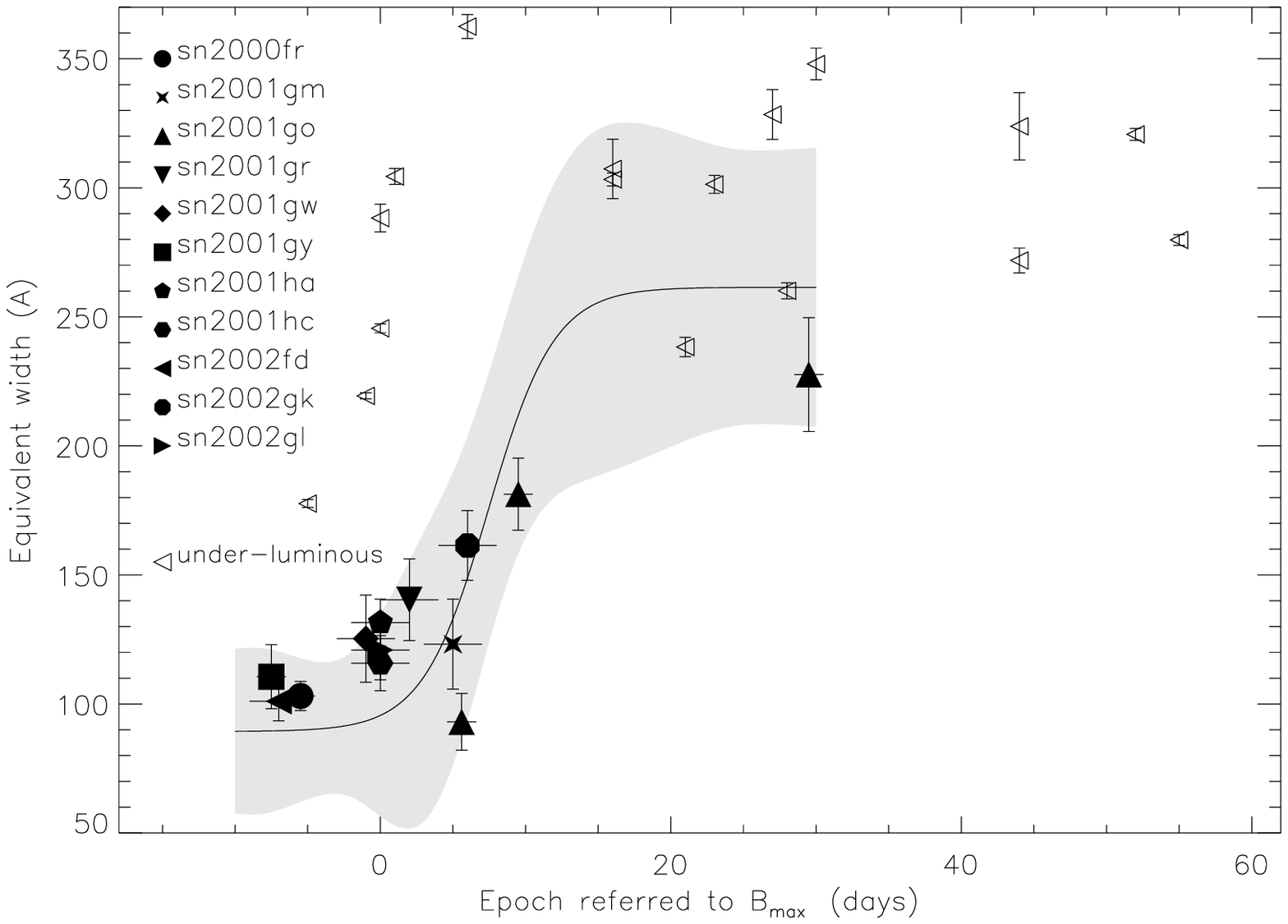}
  \caption{\ \ \ \ \ \ \ \ \ \ \ \ \ \ \ \  \ \ \ \ \ \ \ \ \ \ {\bf``Mg~{\sc ii} 4300''}
  \newline{\it Upper Panel: }Measured pseudo-equivalent widths
    corresponding to ``Mg~{\sc ii} 4300'' (\#3). SN 1991bg-like
    objects are marked with open squares, 1991T-like SNe~Ia with open
    circles, normal SNe~Ia with filled circles and \object{SN 1999ac}
    with triangles. Error-bars include the error
    described by Eqn.~\ref{eq:error} as well as systematic
    uncertainties arising from the pseudo-continuum fit. The
    solid line represents the average behavior of normal SNe~Ia,
    as described in Eqn.~\ref{eq:fd} with  $\vec{\theta}$=(172.2, 89.4, 7.5, 2.5). In general,
    1991bg-like SNe~Ia lie above the average curve whereas 1991T-like
    SNe~Ia lie below it. {\it Lower Panel: } A comparison between the
    `Mg~{\sc ii}~4300' \ew s in low- and high-redshift
    SNe~Ia. High-redshift supernova are indicated by large filled
    symbols. The gray filled area represents the 95\% probability
    region for normal low-redshift SNe~Ia. Peculiar under-luminous nearby SNe~Ia are shown
    separately for comparison. See text for details.}
  \label{fig:MgII}
  \label{fig:Mg_mod}
\end{figure}

Various ions correspond to feature \#3.
These include Mg~{\sc ii}, Co~{\sc
  ii}, Fe~{\sc ii}, Fe~{\sc iii}, and Si~{\sc iii} for
spectroscopically normal and 1991T-like SNe~Ia.  In the case of
1991bg-like SNe~Ia, the region is dominated by strong lines of Ti~{\sc
  ii} \citep{1992AJ....104.1543F,1997MNRAS.284..151M}.  The evolution
of the pseudo-equivalent width of this feature is different than that
of ``Fe~{\sc ii} 4800'' as can be seen in Fig.~\ref{fig:Mg_mod}.

The \ew\ of the feature increases dramatically over a short period of
time as it mergers with the neighboring ``Si~{\sc ii} 4000'' feature
(feature \#3 in figure \ref{fig:Mg_mod}). Before and after this
increase, the \ew\ of this feature is approximately constant. The
phase at which this increase takes place, $t_{br}$, is highly dependent on the
SN~Ia sub-type. For 1991bg-like SNe~Ia it seems to occur as early as 5
days before maximum light (the earliest spectrum of a 1991bg-like SN~Ia
in our sample), while normal SNe~Ia show this behavior around one week
after maximum light, and 1991T-like objects show it later than day
$+10$.  Thus, the evolution of the \ew\ of the ``Mg~{\sc ii}~4300''
feature can be used to discriminate between different Type~Ia
sub-types.

We describe the average evolution of this feature with the function:

\begin{equation}
  \label{eq:fd}
  f(\vec{\theta},t) =
  {A \over e^{{t_{br}-t \over \tau}}+1}+ B,
\end{equation}

with parameters $\vec{\theta} = (A,B,t_{br},\tau)$.  Parameters (A, B)
are simple constants, t is the SN phase, and $\tau$ is  an  e-folding
time scale.
 The solid line in
Fig.~\ref{fig:Mg_mod} shows the average curve for normal SNe~Ia
between days $-10$ and $+30$.  Other SN~Ia subtypes show a parallel
trend.  Table~\ref{tab:MgII_disp} lists the dispersion of normal
SNe~Ia around the average curve and the deviations of 1991T-like and
1991bg-like SNe~Ia from the same curve.  The parameterization
introduced in Eq. \ref{eq:fd} allows one to quantify the
sudden change in the \ew\ in terms of the parameter $t_{br}$, which
can then be used to classify type-Ia supernovae into the three main sub-types.  In
Fig. \ref{fig:Mg_mod} (upper panel) we note that, for normal type~Ia
SNe before maximum light, the \ew\ measurements tend to cluster around high and low
values.  The high values of \ew\ are those measured on SN~1989B (two
data points in the first epoch bin and three in the second one) and on
SN~1990N (one data point in each bin). Instead, SN~1994D shows low \ew
s (two data points in each bin before maximum light), with SN~1992A \ew\ having intermediate values (in both pre-maximum bins). The average
trend is then greatly affected by SN~1994D which, as for feature \#4,
seems to be an outlier.  Because of the low statistics the
pre-maximum trend is only indicative. An analysis
including more normal type~Ia SN will be needed to assess the
average trend and to investigate whether the apparent clustering in
the distribution is a distinctive characteristic of type~Ia SNe
disclosing interesting physics. 

\begin{table*}[htbp]
  \begin{center}
  \caption{Dispersion of ``Mg~{\sc ii} 4300'' \ew\ for the three
  SN~Ia subtypes. See text for details.}  
  \label{tab:MgII_disp}
  \begin{tabular}{l@{\hspace{2.7em}}ccc@{\hspace{2.7em}}cc@{\hspace{2.7em}}cc}
    \hline
    \hline
    Epoch & \multicolumn{3}{c@{\hspace{2.7em}}}{ Normal} &
    \multicolumn{2}{c@{\hspace{2.7em}}}{1991T-like} &
    \multicolumn{2}{c@{\hspace{2.7em}}}{1991bg-like}\\ 
    ~~bin & $n$ & $\langle\mbox{\ew}\rangle$ & 
    $\delta$ & $n$ & 
    $\delta_{\mbox{\scriptsize \ew}}$ & $n$ & 
    $\delta_{\mbox{\scriptsize \ew}}$ \\ 
     & & (\AA) & (\AA) &  & (\AA) &  & (\AA)\\
    $(1)$ & $(2)$ & $(3)$ & $(4)$ & $(5)$ & $(6)$ & $(7)$ & $(8)$\\
    \hline
    $[-10,-5]$ & 6  &  $90\pm7$  & 16 & 6 &  $15\pm4$  & 0 & $\cdots$   \\
    $[-4,0]$   & 6  &  $98\pm5$  & 14 & 5 &  $-5\pm3$  & 4 & $139\pm22$ \\
    $[1,5]$    & 9  & $118\pm12$ & 28 & 4 & $-25\pm12$ & 1 &
    $206$$^{\mathrm{a}}$ \\
    $[6,10]$   & 10 & $184\pm14$ & 26 & 4 & $-77\pm12$ & 0 & $\cdots$   \\
    $[11,17]$  & 15 & $251\pm10$ & 33 & 5 & $-69\pm23$ & 2 &  $48\pm2$  \\
    $[18,30]$  & 15 & $257\pm7$  & 27 & 6 &   $8\pm6$  & 4 &  $21\pm20$ \\
    \hline
  \end{tabular}    
  \end{center}
  \begin{list}{}{}
  \item[] See column explanations in Table~\ref{tab:FeII_disp}.
  \item[$^{\mathrm{a}}$] Only one measurement, thus no uncertainty is
    given.
  \end{list}
\end{table*}

\footnotetext{In
  general we find $\delta\sim\sqrt{N}\cdot\delta_{\mbox{\ew}}$, but
  note that there are slight deviations due to the use of different
  epoch bins for this table and in building the average trend.}

\begin{figure}[htbp]
  \includegraphics[width=\hsize]{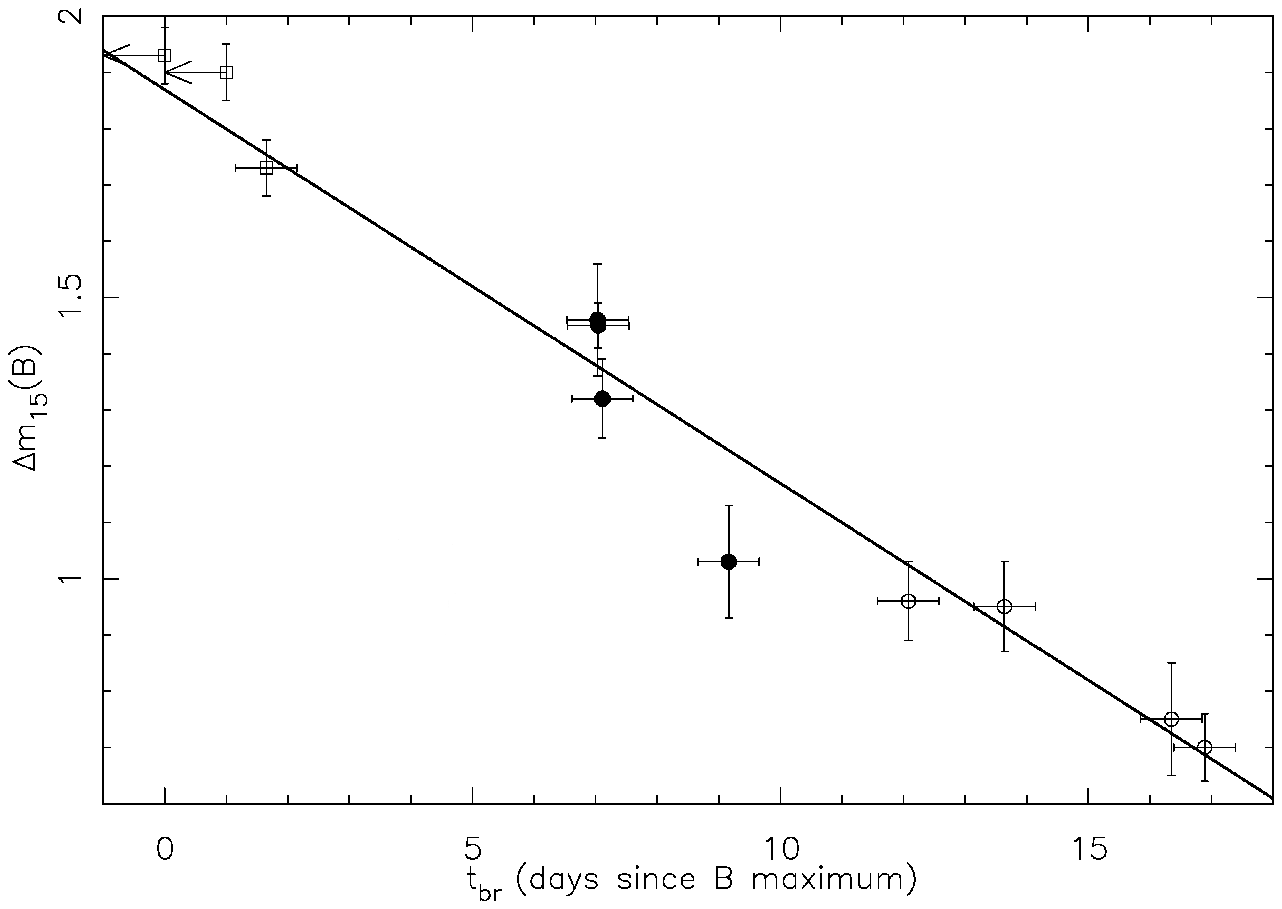}   
  \caption{The ``Mg~{\sc ii} 4300'' \ew\ break parameter $t_{br}$
    (Eqn.~\ref{eq:fd}).  $\Delta m_{15}(B)$ versus $t_{br}$. The two
    points marked with arrows correspond to upper limits of $t_{br}$.
    The straight line is least-squares fit to the data, excluding the
    upper limits.}
  \label{fig:Mgbreak}
\end{figure}

From Fig. \ref{fig:Mg_mod} (upper panel), it is evident that
the evolutionary behavior of the ``Mg~{\sc ii} 4300'' \ew\ 
 is different for normal, 91T-like, and 91bg-like SNe~Ia, and therefore could correlate also with photometric properties of the SN.  To quantify this
correlation, the functional model given by Eq.~\ref{eq:fd} was used to
fit the parameter $t_{br}$ for each of SN~Ia individually. This
parameter is related to the phase at which this feature suddenly
becomes stronger.  Table~\ref{tab:MbDm} lists the values of $t_{br}$
for 11 SNe~Ia in our sample. In the cases of \object{SN 1991bg} and
\object{SN 1999by}, the epochs of their earliest spectra are used as
upper limits for $t_{br}$, assuming they followed the same
evolutionary pattern found for the other objects. The values of
$\Delta m_{15}(B)$ were re-fitted for this analysis to gain
consistency in the results.  Fig.~\ref{fig:Mgbreak} shows the
correlation between $\Delta m_{15}(B)$ and $t_{br}$. The break occurs
at later epochs for SNe~Ia with slower declining lightcurves. A
least-squares fit  yields:

\begin{equation}
\label{eq:t0dm}
\Delta m_{15}(B)=1.869(\pm 0.052)- 0.070(\pm 0.005) t_{br},
\end{equation}

where $1.7<t_{br}<16.9$ is given in days since maximum light. The fit
was done excluding the two upper limits.  The resulting dispersion in
$\Delta m_{15}(B)$ around the fitted line is 0.08 mag, which is
comparable to the measurement uncertainties. Since this spectral
feature is covered by the standard $B$ filter, the correlation is not
surprising. The sudden strengthening of the ``Mg~{\sc ii} 4300''
absorption appears to affect the shape of the lightcurve.

\begin{table*}[htbp]
  \begin{center}
  \caption{$\Delta m_{15}(B)$ and $t_{br}$ measured on the nearby supernova sample.}
  \label{tab:MbDm} 
  \begin{tabular}{lcl}
    \hline
    \hline
    SN & $\Delta m_{15}(B)$$^{\mathrm{a}}$ & $t_{br}$$^{\mathrm{b}}$\\
    \hline
    1986G &  1.73(05) &  1.7 \\  
    1989B & 1.32(07) &  7.1  \\  
    1990N &  1.03(10) &  9.2  \\  
    1991T &  0.96(07) &  12.1  \\ 
    1991bg &  1.93(05) &  $<0$$^{\mathrm{c}}$  \\ 
    1992A &  1.45(04) & 7.0  \\
    1994D &  1.46(10) & 7.0 \\  
    1999aa &  0.95(08) & 13.6  \\ 
    1999aw &  0.75(10) &  16.4  \\  
    1999bp &  0.70(06) & 16.9  \\
    1999by & 1.90(05) & $<1$$^{\mathrm{c}}$  \\ 
    \hline
  \end{tabular}
  \begin{list}{}{}
  \item[$^{\mathrm{a}}$] Uncertainties given between brackets, in
  units of $0.01$ mag. 
  \item[$^{\mathrm{b}}$] Spectral parameter related to 
    ``Mg~{\sc ii} 4300'' (Eqn.~\ref{eq:fd}).  Assumed uncertainty in 
    $t_{br}$: $0.5$d. Measured in days since maximum light.
  \item[$^{\mathrm{c}}$] Upper limit corresponding to the earliest data point.
  \end{list}
  \end{center}
\end{table*}

 An additional point to note is that this feature can be
observed with optical spectrographs up to redshifts of $\sim$ 1, 
and the technique can be used up to redshifts $\geq$ 1.7  with future SN-related
cosmology experiments (e.g ADEPT, DESTINY or SNAP) which will include near-infrared spectrographs. 
However, since several spectroscopic observations are
needed to obtain the value of $t_{br}$, the use of this parameter to
estimate $M_B^{\rm max}$ would be limited to SNe~Ia with rather extensive
spectroscopic follow-up.

Fig~\ref{fig:MgII} (lower panel) shows that the ``Mg~{\sc ii}~4300''
\ew s of the high-redshift supernovae (all except SN~2002gi are
measured) are consistent with the trend defined by the 95\%
probability distribution of low-redshift normal supernova indicated by
the gray filled area. Moreover, SN~2001go, for which measurements at
three epochs are available, shows the sudden increase at around one
week after maximum, as do most low-redshift normal SNe~Ia.

\subsubsection{``Ca~{\sc ii} H\&K'' (\#1)}

\begin{figure}[htbp]
  \includegraphics[width=\hsize]{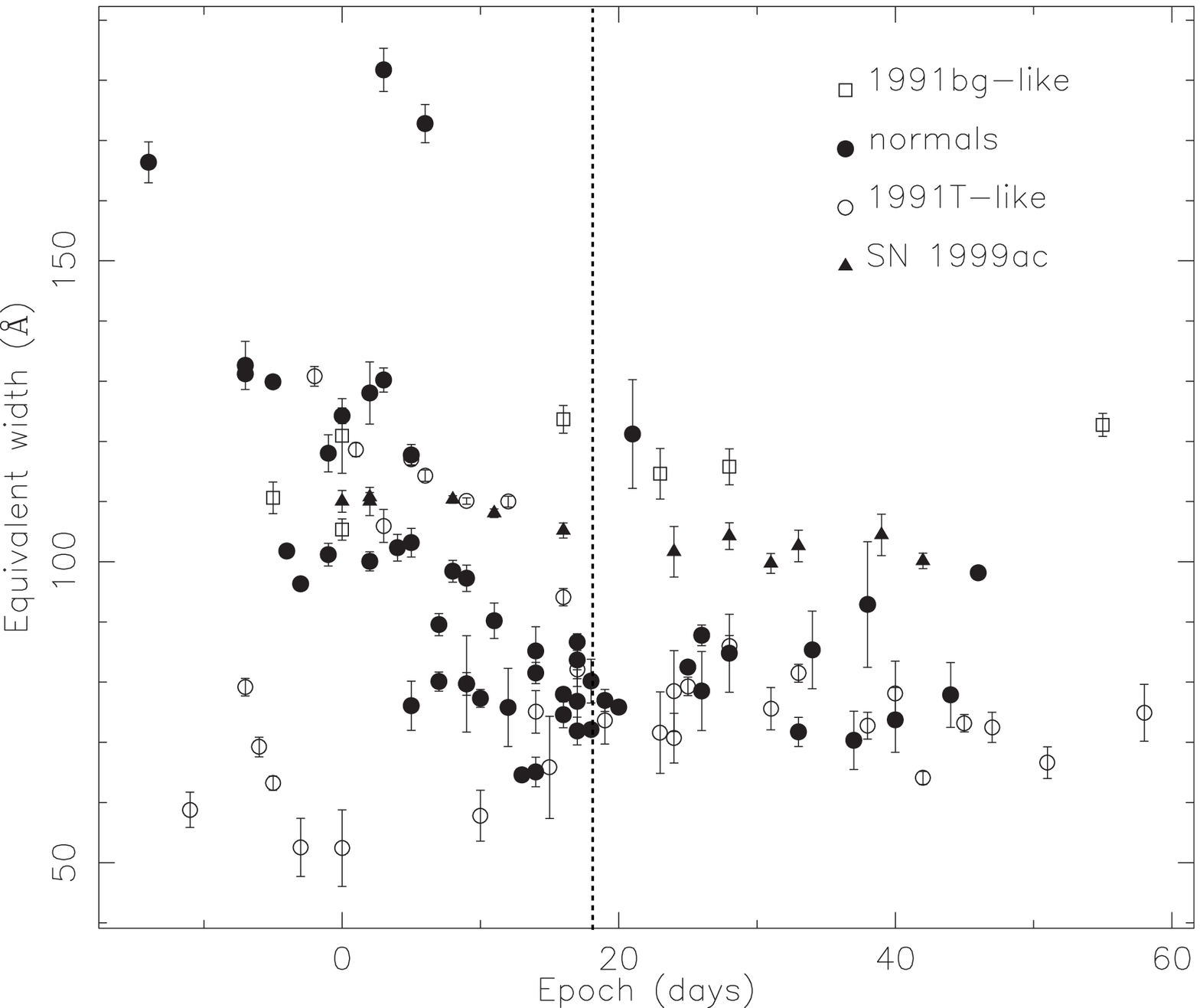}
\centering \includegraphics[width=\hsize]{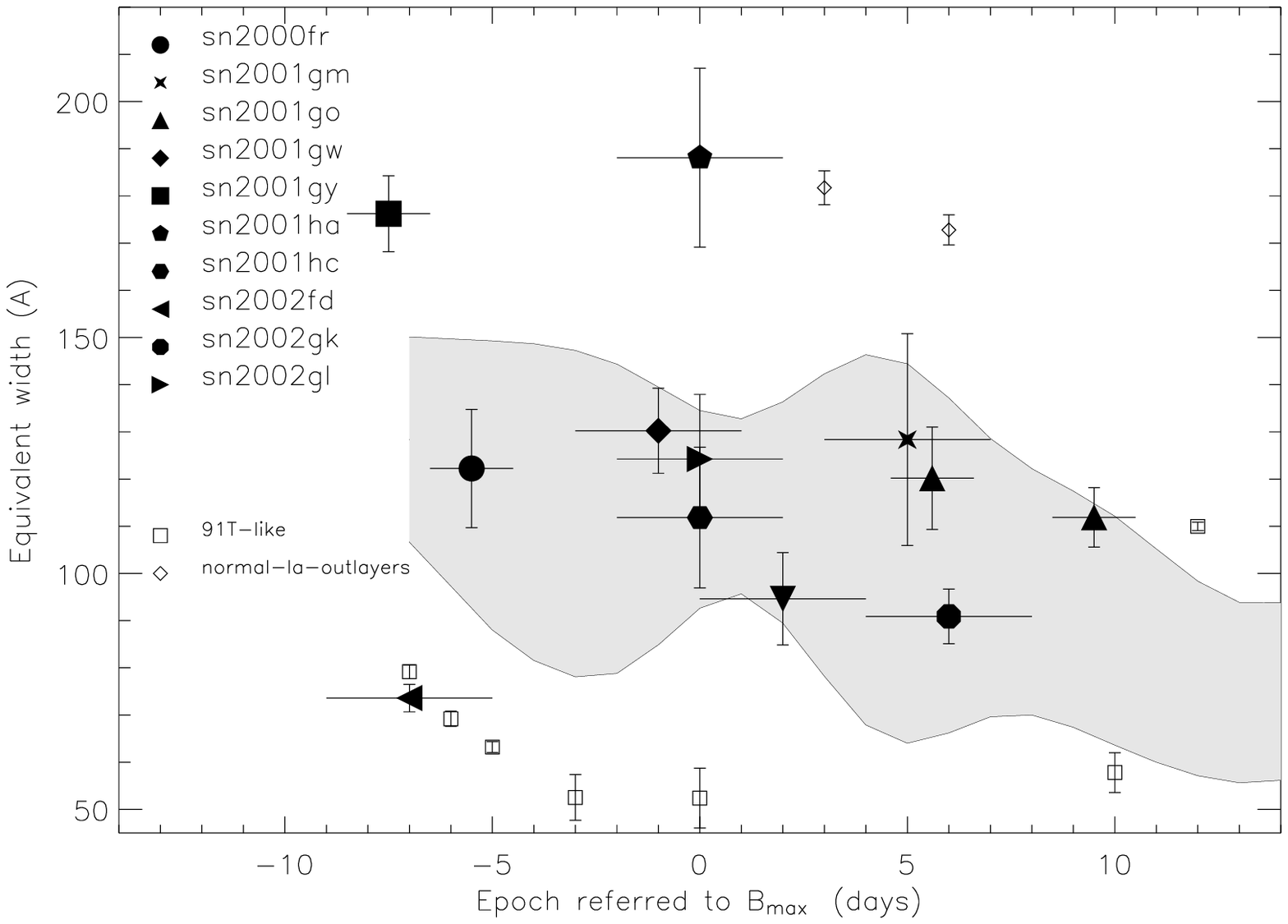}
  \caption{\ \ \ \ \ \ \ \ \ \ \ \ \ \ \ \  \ \ \ \ \ \ \ \ \ \ {\bf``Ca~{\sc ii} H\&K''}
   \newline{\it Upper Panel: }Measured pseudo-equivalent widths
    corresponding to ``Ca~{\sc ii} H\&K'' (\#1). SN 1991bg-like
    objects are marked with open squares, 1991T-like SNe~Ia with open
    circles, normal SNe~Ia with filled circles and \object{SN 1999ac}
    with triangles. Error-bars include the error
    described by Eqn.~\ref{eq:error} as well as systematic
    uncertainties arising from the pseudo-continuum fit.  The vertical dashed line indicates the upper epoch limit in the lower panel.
    {\it Lower Panel: }A comparison between the `Ca~{\sc ii}~H\&K'
    \ew s in low- and high-redshift SNe~Ia. High-redshift
    supernova are indicated by large filled symbols. The gray filled
    area represents the 95\% probability region for normal
    (non-outliers) and the under-luminous low-redshift SNe~Ia
    \object{SN 1999by},\object{SN 1991bg} and \object{SN 1986G}.
    Outliers normal SNe~Ia are indicated with the diamond
    symbols. Slow declining SNe~Ia are indicated by the open square
    symbols.}
  \label{fig:CaHKnearby}
  \label{fig:CaHK}
\end{figure}

The prominent absorption trough at 3800\AA\ is attributed to the
Ca~{\sc ii} H\&K lines with contributions from Si~{\sc ii}
$\lambda$3858 and lines from iron-peak elements.
Fig.~\ref{fig:CaHKnearby} shows the change in the \ew\ of this feature
with phase.  As expected for SN~1991T-like objects, this absorption is
particularly weak (low \ew), especially in the pre-maximum spectra.
The intrinsic dispersion in the \ew s is greater than in the case of
features \#3 (Mg~{\sc ii} 4300) and \#4 (Fe~{\sc ii} 4800). 

This is primarily due to the
strength of the Ca~{\sc ii} ~H\&K line when compared with Mg~{\sc ii} and Fe~{\sc ii} .
The evolution of
the \ew\ with phase is specific to each object both qualitatively and
quantitatively. The dispersion on \ew\ is greater before maximum
light, where some objects show an increase and others a
decrease. After maximum light, there is normally a slow decrease,
except for the 1991bg-like objects in which the \ew\ is relatively
constant. The spread of \ew s of this
feature is the largest of all the features analyzed in this work. This
is due, in part, to the possible presence of high velocity components
and to the overlap of several lines that lie in the wide wavelength
interval spanned by this feature.
As pointed out in \citet{2005ApJ...623L..37M} the position in velocity
space of such components differ in different SNe~Ia making the feature
broader or narrower and thus affecting the measured \ew.

The two points with $\mbox{\ew}>170$\AA\, after day 0 belong to
\object{SN 1999bm}. Their large \ew\ values are due to an unusually
broad ``Ca~{\sc ii} H\&K'' feature.  As pointed out in \citet{2004ApJ...607..391G} this phenomenon might be caused either by the presence of a strong, high-velocity Ca~{\sc ii} component or  by a strong Si~{\sc ii}~3858 line. 
Disentangle between these two hypotheses is difficult. The presence of a high-velocity Ca~{\sc ii} component  could be better determined in the Ca~{\sc ii} IR triplet region, because of the lack of contamination from other ions. However,  in  our spectra of  \object{SN 1999bm}  the signal-to-noise  around 8000 \AA\  is two low to identify such component.  It is impossible to exclude an high-velocity component based only on Ca~{\sc ii}~H\&K.
The strength of the Ca~{\sc ii}~H\&K line could result in a single minimum absorption line 
even when an high-velocity components is present and detected in the Ca II IR triplet.

The \ew\ of ``Ca~{\sc ii}~H\&K'' feature of low- and high-redshift
supernova are shown in Fig.~\ref{fig:CaHK}. For this feature, as
described above, it is not possible to identify a distinctive trend
for SN~Ia subgroups as a function of time. Before twenty days past
maximum, normal and under-luminous low-redshift SNe~Ia populate the
region that spans from \ew\=60 \AA\ to \ew\=140 \AA.  In the lower panel
of Fig.~\ref{fig:CaHK}, the gray filled area represents the 95\%
probability region for normal (non-outliers) and under-luminous
low-redshift SNe~Ia. Some outliers (determined by a 3-sigma clipping algorithm) are found among normal SNe~Ia and
are indicated with small diamond symbols.  Before maximum light, peculiar \taa\
objects show systematically low values, as indicated by the small
square symbols. The high-redshift supernovae (all except SN~2002gi are
plotted) do not show significant deviations with respect to the
low-redshift sample shown in the plot, and SN~2002fd falls on the \taa\
trend as expected.

\subsection{The "Ca~II~H\&K" pseudo-EW of SN~2002fd}
\label{02fdhk}

Prior to maximum light, the \ew\ of Ca~{\sc ii}~H\&K can be used
to separate  \taa\ SN~Ia from normal SNe~Ia, (see
Fig. \ref{fig:CaHK}). If the identification of SN~2002fd as a peculiar
object is correct, we expect the \ew\ to be lower than that of
normal SNe~Ia. The average \ew\ prior to maximum light in
normal SNe~Ia is $<${\mbox{\ew}}$>$ = 114.1 and the scatter around
the mean value is $\sigma_{<{\mbox{\scriptsize {\ew}}}>}$= 14.2 For
peculiar \taa\ SNe~Ia we find $<${\mbox{\ew}}$>$ = 68.7 and
$\sigma_{<{\mbox{\scriptsize \ew}}>}$ = 6.1 The value measured
for SN~2002fd, (${\mbox{\ew}}$ = 73.6 $\pm$ 2.9), is consistent
--- within one standard deviation --- with that found for \taa\
SN~Ia, and inconsistent (at more than 3 standard deviations) with 
normal SNe~Ia.

\label{measurements}

 Note, that the plotted errors bars in the \ew s of the
 high-redshift SNe~Ia include both statistical uncertainties, from the
 measurement, and systematic uncertainties from residual host galaxy
 contamination but do not include systematic uncertainties from
 possible misidentification of the maxima around the absorption
 feature (i.e. misidentification of the fitting regions).  Table
 \ref{ewdata} reports the measured \ew s.

\begin{table*}[htb]
\begin{center}
\small
\caption{Measurements of the \ew s of Ca~{\sc ii}~H\&K, Mg~{\sc ii} and Fe~{\sc ii}. Measurements uncertainties are reported in parenthesis.\tablenotemark{a}}
\begin{tabular}{llllr}
\hline
SN &epoch& Ca~{\sc ii}~H\&K \ew&Mg~{\sc ii} \ew &Fe~{\sc ii} \ew\\
&[days]& [\AA]&[\AA]&[\AA]\\
\hline
sn2001gy& $-$7.5 (1) &  176 (8/21)& 111 (12/31)&    \\
sn2002fd& $-$7 (2)&  73.6 (3/14) &101 (8/11)  &    86 (15/23)\\
sn2000fr& $-$5.5 (1) & 122 (13/14)& 103 (6/13) &   91 (13/23)\\
sn2001gw& $-$1 (2) & 130 (9/13) & 125 (17/22) &        \\
sn2002gl& 0 (2)&  124 (14/15)&  121 (12/18)&      168 (20/25)\\
sn2001ha& 0 (2) & 188 (19/22) & 132 (9/24) &   216 (33/62)\\
sn2001hc& 0 (2) & 112 (15/15)& 116 (11/18)     &   140 (15/23)\\ 
sn2001gr& 2 (2)&  94.6 (10/22)& 140 (16/34) &   141 (31/69)\\
sn2001gm& 5 (2)&  128 (23/26) &123 (17/40)&        169 (34/84)\\
sn2001go& 5.6 (1)&    120 (11/28)&    93 (11/24)    &     191 (28/41)\\
sn2002gk& 6 (2) &91 (6/11)& 161 (14/16) &157 (16/17)     \\
sn2001go& 9.5 (1) &   112 (6/13) & 181 (14/24) &  174 (20/27)\\
sn2001go& 29.5 (1)   & 228 (22/48)  &      315 (41/52)\\
\hline 
\end{tabular}
\tablenotetext{a}{The first value includes statistical and systematic uncertainties
related to the level of host galaxy contamination, the second value (added in
  quadrature) also includes possible systematic uncertainties due to
  misidentification of the fitting regions, (see sec. \ref{ew} for
  details).}
\label{ewdata}
\end{center}
\end{table*}

\subsection{Statistical comparison}
\label{stat}

In this section, a statistical comparison of the
low- and high-redshift SNe~Ia is performed using the spectral indicators
described in section \ref{ew}.  The mean trends in the \ew s of
the `Fe~{\sc ii} 4800' and `Mg~{\sc ii} 4300' features identified for
normal -- low-redshift -- SNe~Ia can be used to test whether or not
high-redshift supernovae \ew s follow the same trends.

The results of a set of $\chi^{2}$ tests are shown in Table
\ref{emdata}. The intrinsic dispersion around the fitted mean trends
for normal low-$z$ supernovae was added in quadrature to the
statistical and systematic uncertainties to perform the test. The
uncertainty in the SN~Ia phase was propagated according to the \ew\
model for normal low redshift SNe~Ia and added in quadrature to the
measurements error on the \ew s.  
The possible systematic uncertainties due to misidentification of the maxima was
included was also taken into consideration. We note that the fitting region uncertainties --- included
in the results shown in Table \ref{emdata} --- should be considered as upper limits to the
possible systematic uncertainty due to the \ew 's measurement
technique (see section \ref{ew} for details).

The hypothesis that the \ew\ measured on our
high-redshift supernovae follow the same behavior with lightcurve
phase as those measured on low redshift normal supernovae is
statistically confirmed. Moreover, the hypothesis that \ew s
measured on our high-redshift supernovae are consistent with those of
under-luminous SN~1991bg-like low redshift SNe~Ia is rejected.

\begin{table*}[htb]
\begin{center}
\small
\caption{ A statistical comparison of the `Fe~{\sc ii} 4800' and
  `Mg~{\sc ii} 4300' pseudo-equivalent widths of high and low-redshift
  SNe~Ia.}
  
\begin{tabular}{|c|cccc|cccc|}
\hline
Feature&$\chi^{2}_{\rm norm}$ &P$_{\rm norm}$&$\chi^{2}_{\rm 91bg-like}$ &P$_{\rm 91bg-like}$&$\chi^{2}_{\rm norm}$ &P$_{\rm norm}$&$\chi^{2}_{\rm 91bg-like}$ &P$_{\rm 91bg-like}$ \\
 &{\tiny (2)}&{\tiny(3)}&{\tiny(4)}&{\tiny(5)}&{\tiny(6)}&{\tiny(7)}&{\tiny(8)}&{\tiny(9)}\\ 
\hline 

 Fe~{\sc ii}&13.0(11)&0.29&29.6(11)& 0.002&6.7(11)&0.8&16.8(11)& 0.1\\ 
 Mg~{\sc ii}&11.2(13)&0.59& 228.4(13)&0.00&7.3(13)&0.9& 162.4(13)&0.00\\
\hline
\end{tabular}
\label{emdata}
  \begin{list}{}{}
  \item[]
  Columns 2, 3, 6 and 7 report the $\chi^{2}$
correspond to the comparison of the \ew s `Fe~{\sc ii} 4800' and
`Mg~{\sc ii} 4300' of high redshift SNe~Ia with the fits for normal
supernovae. Columns 4, 5, 8 and 9 report  the comparison
with under-luminous SNe~Ia (e.g. SN~1986G and SN~1991bg).  Columns 2,
4, 6,and 8 reports the $\chi^{2}$ obtained with the number of degrees
of freedom in parenthesis. Columns 3, 5, 7, and 9 report the
probability to obtain a $\chi^{2}$ value greater than that
obtained. Columns 2 to 6 report the results including both statistical
uncertainties and systematic uncertainties due to host galaxy
subtraction. Columns 7 to 9 report the results when the
systematic uncertainty due to possible misidentification of fitting
regions is also included. 
  \end{list}

\end{center}
\end{table*}

\section{Summary and Conclusions}
\label{conc}

Spectroscopic data of 12 high-redshift supernovae, in the redshift
interval $z$=0.212 to 0.912, were analyzed and a qualitative
classification scheme was presented.  Based on this classification
scheme, all of our high-redshift SNe~Ia were classified as normal
SNe~Ia, except for SN~2002fd ($z$=0.27), which is similar to
SN~1999aa, a peculiar \taa\ SN~Ia.  We also find, based on spectral
properties alone, that none of the supernovae studied in this work are
under-luminous. This is not unexpected because of the bias against
selecting such objects in magnitude limited surveys (see for example
\citet{2001ApJ...546..719L}).

A quantitative comparison between low and high-redshift SNe~Ia by
means of spectral indicators has been presented. The velocities of the minimum of
Ca~{\sc ii}~H\&K feature of high-redshift \sne, were compared to those of low-redshift
SNe~Ia with the aim of uncovering differences. No systematic
differences could be found.

Using a low-redshift SN Ia sample, we study the evolution in the
\ew\ of the strongest spectral features as a function of phase, and
find that the \ew\ of different SNe~Ia sub-types (normal, 91T-like and
91bg-like) evolve differently for two of the three features studied
(i.e. ``Fe~{\sc ii} 4800'' and ``Mg~{\sc ii} 4300'').  In the case of
``Fe~{\sc ii} 4800'', the different SNe~Ia sub-types follow similar trends, but offset from the average evolution.
91bg-like SNe~Ia show higher \ew s and
91T-like SNe~Ia lower \ew s than normal SNe~Ia. The \ew\ of ``Mg~{\sc
  ii} 4300'' is characterized by a sudden break around maximum
light. The epoch at which the break occurs correlates with the
photometric properties of the SNe~Ia.  We find that in 91bg-like
SNe~Ia the break occurs earlier in phase with respect to normal
SNe~Ia while in 91T-like SNe~Ia the same occurs later.

The pseudo-equivalent widths of ``Fe~{\sc ii} 4800'', ``Mg~{\sc ii} 4300''
and ``Ca~{\sc ii}~H\&K'' in high-redshift SNe~Ia are found to follow
the same trends with epoch as those observed in normal low-redshift
\sne.  Furthermore, the pseudo-equivalent widths of ``Fe~{\sc ii} 4800'' and
``Mg~{\sc ii} 4300'' in high-redshift SNe~Ia are found to be
statistically consistent with the pseudo-equivalent widths observed in
low-redshift normal SNe~Ia.

The \ew s of Ca~{\sc ii}~H\&K in the spectrum of SN~2002fd are
consistent with those observed in SN~1991T/SN1999aa-like objects in
the local Universe, quantitatively confirming the sub-type
identification of this SN~Ia. This
feature can be used to identify \taa\ objects at high-redshift.

The number of higher-redshift supernova used in this case study is small when
compared to the numbers of SNe~Ia that are now being observed  at similar redshifts.  Both
the ESSENCE and SNLS projects will result in samples of SNe~Ia with
similar spectral quality; however, the samples will be many times
larger. 
The case study presented here
offers a simple method to analyze the spectra observed in these
surveys and look for systematic differences between SNe Ia at
different redshift.


\bibliography{Evolution_v5.2.bbl}

\begin{thebibliography}{61}
\expandafter\ifx\csname natexlab\endcsname\relax\def\natexlab#1{#1}\fi

\bibitem[{{Astier} {et~al.}(2006){Astier}, {Guy}, {Regnault}, {Pain},
  {Aubourg}, {Balam}, {Basa}, {Carlberg}, {Fabbro}, {Fouchez}, {Hook},
  {Howell}, {Lafoux}, {Neill}, {Palanque-Delabrouille}, {Perrett}, {Pritchet},
  {Rich}, {Sullivan}, {Taillet}, {Aldering}, {Antilogus}, {Arsenijevic},
  {Balland}, {Baumont}, {Bronder}, {Courtois}, {Ellis}, {Filiol}, {Gon{\c
  c}alves}, {Goobar}, {Guide}, {Hardin}, {Lusset}, {Lidman}, {McMahon},
  {Mouchet}, {Mourao}, {Perlmutter}, {Ripoche}, {Tao}, \&
  {Walton}}]{2006A&A...447...31A}
{Astier}, P., {Guy}, J., {Regnault}, N., {et~al.} 2006, \aap, 447, 31

\bibitem[{{Balland} {et~al.}(2006){Balland}, {Mouchet}, {Pain}, {Walton},
  {Amanullah}, {Astier}, {Ellis}, {Fabbro}, {Goobar}, {Hardin}, {Hook},
  {Irwin}, {McMahon}, {Mendez}, {Ruiz-Lapuente}, {Sainton}, {Schahmaneche}, \&
  {Stanishev}}]{2006A&A...445..387B}
{Balland}, C., {Mouchet}, M., {Pain}, R., {et~al.} 2006, \aap, 445, 387

\bibitem[{{Barris} {et~al.}(2004){Barris}, {Tonry}, {Blondin}, {Challis},
  {Chornock}, {Clocchiatti}, {Filippenko}, {Garnavich}, {Holland}, {Jha},
  {Kirshner}, {Krisciunas}, {Leibundgut}, {Li}, {Matheson}, {Miknaitis},
  {Riess}, {Schmidt}, {Smith}, {Sollerman}, {Spyromilio}, {Stubbs}, {Suntzeff},
  {Aussel}, {Chambers}, {Connelley}, {Donovan}, {Henry}, {Kaiser}, {Liu},
  {Mart{\'{\i}}n}, \& {Wainscoat}}]{2004ApJ...602..571B}
{Barris}, B.~J., {Tonry}, J.~L., {Blondin}, S., {et~al.} 2004, \apj, 602, 571

\bibitem[{{Blakeslee} {et~al.}(2003){Blakeslee}, {Tsvetanov}, {Riess}, {Ford},
  {Illingworth}, {Magee}, {Tonry}, {Ben{\'{\i}}tez}, {Clampin}, {Hartig},
  {Meurer}, {Sirianni}, {Ardila}, {Bartko}, {Bouwens}, {Broadhurst}, {Cross},
  {Feldman}, {Franx}, {Golimowski}, {Gronwall}, {Kimble}, {Krist}, {Martel},
  {Menanteau}, {Miley}, {Postman}, {Rosati}, {Sparks}, {Strolger}, {Tran},
  {White}, \& {Zheng}}]{2003ApJ...589..693B}
{Blakeslee}, J.~P., {Tsvetanov}, Z.~I., {Riess}, A.~G., {et~al.} 2003, \apj,
  589, 693

\bibitem[{{Blondin} {et~al.}(2006){Blondin}, {Dessart}, {Leibundgut}, {Branch},
  {H{\"o}flich}, {Tonry}, {Matheson}, {Foley}, {Chornock}, {Filippenko},
  {Sollerman}, {Spyromilio}, {Kirshner}, {Wood-Vasey}, {Clocchiatti},
  {Aguilera}, {Barris}, {Becker}, {Challis}, {Covarrubias}, {Davis},
  {Garnavich}, {Hicken}, {Jha}, {Krisciunas}, {Li}, {Miceli}, {Miknaitis},
  {Pignata}, {Prieto}, {Rest}, {Riess}, {Salvo}, {Schmidt}, {Smith}, {Stubbs},
  \& {Suntzeff}}]{2006AJ....131.1648B}
{Blondin}, S., {Dessart}, L., {Leibundgut}, B., {et~al.} 2006, \aj, 131, 1648

\bibitem[{{Branch} {et~al.}(2006){Branch}, {Dang}, {Hall}, {Ketchum},
  {Melakayil}, {Parrent}, {Troxel}, {Casebeer}, {Jeffery}, \&
  {Baron}}]{2006PASP..118..560B}
{Branch}, D., {Dang}, L.~C., {Hall}, N., {et~al.} 2006, \pasp, 118, 560

\bibitem[{{Branch} {et~al.}(1983){Branch}, {Lacy}, {McCall}, {Sutherland},
  {Uomoto}, {Wheeler}, \& {Wills}}]{1983ApJ...270..123B}
{Branch}, D., {Lacy}, C.~H., {McCall}, M.~L., {et~al.} 1983, \apj, 270, 123

\bibitem[{{Cardelli} {et~al.}(1989){Cardelli}, {Clayton}, \&
  {Mathis}}]{1989ApJ...345..245C}
{Cardelli}, J.~A., {Clayton}, G.~C., \& {Mathis}, J.~S. 1989, \apj, 345, 245

\bibitem[{{Coil} {et~al.}(2000){Coil}, {Matheson}, {Filippenko}, {Leonard},
  {Tonry}, {Riess}, {Challis}, {Clocchiatti}, {Garnavich}, {Hogan}, {Jha},
  {Kirshner}, {Leibundgut}, {Phillips}, {Schmidt}, {Schommer}, {Smith},
  {Soderberg}, {Spyromilio}, {Stubbs}, {Suntzeff}, \&
  {Woudt}}]{2000ApJ...544L.111C}
{Coil}, A.~L., {Matheson}, T., {Filippenko}, A.~V., {et~al.} 2000, \apjl, 544,
  L111

\bibitem[{{Cristiani} {et~al.}(1992){Cristiani}, {Cappellaro}, {Turatto},
  {Bergeron}, {Bues}, {Buson}, {Danziger}, {di Serego-Alighieri}, {Duerbeck},
  {Heydari-Malayeri}, {Krautter}, {Schmutz}, \&
  {Schulte-Ladbeck}}]{1992A&A...259...63C}
{Cristiani}, S., {Cappellaro}, E., {Turatto}, M., {et~al.} 1992, \aap, 259, 63

\bibitem[{{Filippenko}(1997)}]{1997ARA&A..35..309F}
{Filippenko}, A.~V. 1997, \araa, 35, 309

\bibitem[{{Filippenko} {et~al.}(1992{\natexlab{a}}){Filippenko}, {Richmond},
  {Branch}, {Gaskell}, {Herbst}, {Ford}, {Treffers}, {Matheson}, {Ho}, {Dey},
  {Sargent}, {Small}, \& {van Breugel}}]{1992AJ....104.1543F}
{Filippenko}, A.~V., {Richmond}, M.~W., {Branch}, D., {et~al.}
  1992{\natexlab{a}}, \aj, 104, 1543

\bibitem[{{Filippenko} {et~al.}(1992{\natexlab{b}}){Filippenko}, {Richmond},
  {Matheson}, {Shields}, {Burbidge}, {Cohen}, {Dickinson}, {Malkan}, {Nelson},
  {Pietz}, {Schlegel}, {Schmeer}, {Spinrad}, {Steidel}, {Tran}, \&
  {Wren}}]{1992ApJ...384L..15F}
{Filippenko}, A.~V., {Richmond}, M.~W., {Matheson}, T., {et~al.}
  1992{\natexlab{b}}, \apjl, 384, L15

\bibitem[{{Folatelli}(2004)}]{gastonthesis}
{Folatelli}, G. 2004, Ph.D.~Thesis, available at http://www.cops.physto.se/

\bibitem[{{Gal-Yam} {et~al.}(1999){Gal-Yam}, {Maoz}, {Strolger}, {Smith},
  {Goobar}, {Dahlen}, {Hook}, {Nugent}, {Phillips}, \&
  {Lidman}}]{1999IAUC.7130....1G}
{Gal-Yam}, A., {Maoz}, D., {Strolger}, L.~G., {et~al.} 1999, in International
  Astronomical Union Circular, 1--+

\bibitem[{{Garavini} {et~al.}(2005){Garavini}, {Aldering}, {Amadon},
  {Amanullah}, {Astier}, {Balland}, {Blanc}, {Conley}, {Dahl{\'e}n}, {Deustua},
  {Ellis}, {Fabbro}, {Fadeyev}, {Fan}, {Folatelli}, {Frye}, {Gates}, {Gibbons},
  {Goldhaber}, {Goldman}, {Goobar}, {Groom}, {Haissinski}, {Hardin}, {Hook},
  {Howell}, {Kent}, {Kim}, {Knop}, {Kowalski}, {Kuznetsova}, {Lee}, {Lidman},
  {Mendez}, {Miller}, {Moniez}, {Mouchet}, {Mour{\~a}o}, {Newberg}, {Nobili},
  {Nugent}, {Pain}, {Perdereau}, {Perlmutter}, {Quimby}, {Regnault}, {Rich},
  {Richards}, {Ruiz-Lapuente}, {Schaefer}, {Schahmaneche}, {Smith},
  {Spadafora}, {Stanishev}, {Thomas}, {Walton}, {Wang}, \&
  {Wood-Vasey}}]{2005AJ....130.2278G}
{Garavini}, G., {Aldering}, G., {Amadon}, A., {et~al.} 2005, \aj, 130, 2278

\bibitem[{{Garavini} {et~al.}(2004){Garavini}, {Folatelli}, {Goobar}, {Nobili},
  {Aldering}, {Amadon}, {Amanullah}, {Astier}, {Balland}, {Blanc}, {Burns},
  {Conley}, {Dahl{\'e}n}, {Deustua}, {Ellis}, {Fabbro}, {Fan}, {Frye}, {Gates},
  {Gibbons}, {Goldhaber}, {Goldman}, {Groom}, {Haissinski}, {Hardin}, {Hook},
  {Howell}, {Kasen}, {Kent}, {Kim}, {Knop}, {Lee}, {Lidman}, {Mendez},
  {Miller}, {Moniez}, {Mour{\~a}o}, {Newberg}, {Nugent}, {Pain}, {Perdereau},
  {Perlmutter}, {Prasad}, {Quimby}, {Raux}, {Regnault}, {Rich}, {Richards},
  {Ruiz-Lapuente}, {Sainton}, {Schaefer}, {Schahmaneche}, {Smith}, {Spadafora},
  {Stanishev}, {Walton}, {Wang}, \& {Wood-Vasey}}]{2004AJ....128..387G}
{Garavini}, G., {Folatelli}, G., {Goobar}, A., {et~al.} 2004, \aj, 128, 387

\bibitem[{{Garnavich} {et~al.}(2004){Garnavich}, {Bonanos}, {Krisciunas},
  {Jha}, {Kirshner}, {Schlegel}, {Challis}, {Macri}, {Hatano}, {Branch},
  {Bothun}, \& {Freedman}}]{2004ApJ...613.1120G}
{Garnavich}, P.~M., {Bonanos}, A.~Z., {Krisciunas}, K., {et~al.} 2004, \apj,
  613, 1120

\bibitem[{{Garnavich} {et~al.}(1998){Garnavich}, {Kirshner}, {Challis},
  {Tonry}, {Gilliland}, {Smith}, {Clocchiatti}, {Diercks}, {Filippenko},
  {Hamuy}, {Hogan}, {Leibundgut}, {Phillips}, {Reiss}, {Riess}, {Schmidt},
  {Schommer}, {Spyromilio}, {Stubbs}, {Suntzeff}, \&
  {Wells}}]{1998ApJ...493L..53G}
{Garnavich}, P.~M., {Kirshner}, R.~P., {Challis}, P., {et~al.} 1998, \apjl,
  493, L53+

\bibitem[{{Gerardy} {et~al.}(2004){Gerardy}, {H{\"o}flich}, {Fesen}, {Marion},
  {Nomoto}, {Quimby}, {Schaefer}, {Wang}, \& {Wheeler}}]{2004ApJ...607..391G}
{Gerardy}, C.~L., {H{\"o}flich}, P., {Fesen}, R.~A., {et~al.} 2004, \apj, 607,
  391

\bibitem[{{Goldhaber} {et~al.}(2001){Goldhaber}, {Groom}, {Kim}, {Aldering},
  {Astier}, {Conley}, {Deustua}, {Ellis}, {Fabbro}, {Fruchter}, {Goobar},
  {Hook}, {Irwin}, {Kim}, {Knop}, {Lidman}, {McMahon}, {Nugent}, {Pain},
  {Panagia}, {Pennypacker}, {Perlmutter}, {Ruiz-Lapuente}, {Schaefer},
  {Walton}, \& {York}}]{2001ApJ...558..359G}
{Goldhaber}, G., {Groom}, D.~E., {Kim}, A., {et~al.} 2001, \apj, 558, 359

\bibitem[{{Gray}(1992)}]{1992Sci...257.1978G}
{Gray}, D.~F. 1992, Science, 257, 1978

\bibitem[{{Hamuy} {et~al.}(2002){Hamuy}, {Maza}, {Pinto}, {Phillips},
  {Suntzeff}, {Blum}, {Olsen}, {Pinfield}, {Ivanov}, {Augusteijn}, {Brillant},
  {Chadid}, {Cuby}, {Doublier}, {Hainaut}, {Le Floc'h}, {Lidman},
  {Petr-Gotzens}, {Pompei}, \& {Vanzi}}]{2002AJ....124..417H}
{Hamuy}, M., {Maza}, J., {Pinto}, P.~A., {et~al.} 2002, \aj, 124, 417

\bibitem[{{Hardin} {et~al.}(2000){Hardin}, {Afonso}, {Alard}, {Albert},
  {Amadon}, {Andersen}, {Ansari}, {Aubourg}, {Bareyre}, {Bauer}, {Beaulieu},
  {Blanc}, {Bouquet}, {Char}, {Charlot}, {Couchot}, {Coutures}, {Derue},
  {Ferlet}, {Glicenstein}, {Goldman}, {Gould}, {Graff}, {Gros}, {Haissinski},
  {Hamilton}, {de Kat}, {Kim}, {Lasserre}, {Lesquoy}, {Loup}, {Magneville},
  {Mansoux}, {Marquette}, {Maurice}, {Milsztajn}, {Moniez},
  {Palanque-Delabrouille}, {Perdereau}, {Pr{\'e}vot}, {Regnault}, {Rich},
  {Spiro}, {Vidal-Madjar}, {Vigroux}, {Zylberajch}, \& {The EROS
  Collaboration}}]{2000A&A...362..419H}
{Hardin}, D., {Afonso}, C., {Alard}, C., {et~al.} 2000, \aap, 362, 419

\bibitem[{{Hoeflich} {et~al.}(1998){Hoeflich}, {Wheeler}, \&
  {Thielemann}}]{1998ApJ...495..617H}
{Hoeflich}, P., {Wheeler}, J.~C., \& {Thielemann}, F.~K. 1998, \apj, 495, 617

\bibitem[{{Hook} {et~al.}(2005){Hook}, {Howell}, {Aldering}, {Amanullah},
  {Burns}, {Conley}, {Deustua}, {Ellis}, {Fabbro}, {Fadeyev}, {Folatelli},
  {Garavini}, {Gibbons}, {Goldhaber}, {Goobar}, {Groom}, {Kim}, {Knop},
  {Kowalski}, {Lidman}, {Nobili}, {Nugent}, {Pain}, {Pennypacker},
  {Perlmutter}, {Ruiz-Lapuente}, {Sainton}, {Schaefer}, {Smith}, {Spadafora},
  {Stanishev}, {Thomas}, {Walton}, {Wang}, \&
  {Wood-Vasey}}]{2005AJ....130.2788H}
{Hook}, I.~M., {Howell}, D.~A., {Aldering}, G., {et~al.} 2005, \aj, 130, 2788

\bibitem[{{Howell} {et~al.}(2005){Howell}, {Sullivan}, {Perrett}, {Bronder},
  {Hook}, {Astier}, {Aubourg}, {Balam}, {Basa}, {Carlberg}, {Fabbro},
  {Fouchez}, {Guy}, {Lafoux}, {Neill}, {Pain}, {Palanque-Delabrouille},
  {Pritchet}, {Regnault}, {Rich}, {Taillet}, {Knop}, {McMahon}, {Perlmutter},
  \& {Walton}}]{2005ApJ...634.1190H}
{Howell}, D.~A., {Sullivan}, M., {Perrett}, K., {et~al.} 2005, \apj, 634, 1190

\bibitem[{{Jeffery} {et~al.}(1992){Jeffery}, {Leibundgut}, {Kirshner},
  {Benetti}, {Branch}, \& {Sonneborn}}]{1992ApJ...397..304J}
{Jeffery}, D.~J., {Leibundgut}, B., {Kirshner}, R.~P., {et~al.} 1992, \apj,
  397, 304

\bibitem[{{Kirshner} {et~al.}(1993){Kirshner}, {Jeffery}, {Leibundgut},
  {Challis}, {Sonneborn}, {Phillips}, {Suntzeff}, {Smith}, {Winkler}, {Winge},
  {Hamuy}, {Hunter}, {Roth}, {Blades}, {Branch}, {Chevalier}, {Fransson},
  {Panagia}, {Wagoner}, {Wheeler}, \& {Harkness}}]{1993ApJ...415..589K}
{Kirshner}, R.~P., {Jeffery}, D.~J., {Leibundgut}, B., {et~al.} 1993, \apj,
  415, 589

\bibitem[{{Knop} {et~al.}(2003){Knop}, {Aldering}, {Amanullah}, {Astier},
  {Blanc}, {Burns}, {Conley}, {Deustua}, {Doi}, {Ellis}, {Fabbro}, {Folatelli},
  {Fruchter}, {Garavini}, {Garmond}, {Garton}, {Gibbons}, {Goldhaber},
  {Goobar}, {Groom}, {Hardin}, {Hook}, {Howell}, {Kim}, {Lee}, {Lidman},
  {Mendez}, {Nobili}, {Nugent}, {Pain}, {Panagia}, {Pennypacker}, {Perlmutter},
  {Quimby}, {Raux}, {Regnault}, {Ruiz-Lapuente}, {Sainton}, {Schaefer},
  {Schahmaneche}, {Smith}, {Spadafora}, {Stanishev}, {Sullivan}, {Walton},
  {Wang}, {Wood-Vasey}, \& {Yasuda}}]{2003ApJ...598..102K}
{Knop}, R.~A., {Aldering}, G., {Amanullah}, R., {et~al.} 2003, \apj, 598, 102

\bibitem[{{Krisciunas} {et~al.}(2005){Krisciunas}, {Garnavich}, {Challis},
  {Prieto}, {Riess}, {Barris}, {Aguilera}, {Becker}, {Blondin}, {Chornock},
  {Clocchiatti}, {Covarrubias}, {Filippenko}, {Foley}, {Hicken}, {Jha},
  {Kirshner}, {Leibundgut}, {Li}, {Matheson}, {Miceli}, {Miknaitis}, {Rest},
  {Salvo}, {Schmidt}, {Smith}, {Sollerman}, {Spyromilio}, {Stubbs}, {Suntzeff},
  {Tonry}, \& {Wood-Vasey}}]{2005AJ....130.2453K}
{Krisciunas}, K., {Garnavich}, P.~M., {Challis}, P., {et~al.} 2005, \aj, 130,
  2453

\bibitem[{{Leibundgut} {et~al.}(1991){Leibundgut}, {Kirshner}, {Filippenko},
  {Shields}, {Foltz}, {Phillips}, \& {Sonneborn}}]{1991ApJ...371L..23L}
{Leibundgut}, B., {Kirshner}, R.~P., {Filippenko}, A.~V., {et~al.} 1991, \apjl,
  371, L23

\bibitem[{{Leibundgut} {et~al.}(1993){Leibundgut}, {Kirshner}, {Phillips},
  {Wells}, {Suntzeff}, {Hamuy}, {Schommer}, {Walker}, {Gonzalez}, {Ugarte},
  {Williams}, {Williger}, {Gomez}, {Marzke}, {Schmidt}, {Whitney}, {Coldwell},
  {Peters}, {Chaffee}, {Foltz}, {Rehner}, {Siciliano}, {Barnes}, {Cheng},
  {Hintzen}, {Kim}, {Maza}, {Parker}, {Porter}, {Schmidtke}, \&
  {Sonneborn}}]{1993AJ....105..301L}
{Leibundgut}, B., {Kirshner}, R.~P., {Phillips}, M.~M., {et~al.} 1993, \aj,
  105, 301

\bibitem[{{Lentz} {et~al.}(2000){Lentz}, {Baron}, {Branch}, {Hauschildt}, \&
  {Nugent}}]{2000ApJ...530..966L}
{Lentz}, E.~J., {Baron}, E., {Branch}, D., {Hauschildt}, P.~H., \& {Nugent},
  P.~E. 2000, \apj, 530, 966

\bibitem[{{Li} {et~al.}(2003){Li}, {Filippenko}, {Chornock}, {Berger},
  {Berlind}, {Calkins}, {Challis}, {Fassnacht}, {Jha}, {Kirshner}, {Matheson},
  {Sargent}, {Simcoe}, {Smith}, \& {Squires}}]{2003PASP..115..453L}
{Li}, W., {Filippenko}, A.~V., {Chornock}, R., {et~al.} 2003, \pasp, 115, 453

\bibitem[{{Li} {et~al.}(2001{\natexlab{a}}){Li}, {Filippenko}, \&
  {Riess}}]{2001ApJ...546..719L}
{Li}, W., {Filippenko}, A.~V., \& {Riess}, A.~G. 2001{\natexlab{a}}, \apj, 546,
  719

\bibitem[{{Li} {et~al.}(2001{\natexlab{b}}){Li}, {Filippenko}, {Treffers},
  {Riess}, {Hu}, \& {Qiu}}]{2001ApJ...546..734L}
{Li}, W., {Filippenko}, A.~V., {Treffers}, R.~R., {et~al.} 2001{\natexlab{b}},
  \apj, 546, 734

\bibitem[{{Lidman} {et~al.}(2005){Lidman}, {Howell}, {Folatelli}, {Garavini},
  {Nobili}, {Aldering}, {Amanullah}, {Antilogus}, {Astier}, {Blanc}, {Burns},
  {Conley}, {Deustua}, {Doi}, {Ellis}, {Fabbro}, {Fadeyev}, {Gibbons},
  {Goldhaber}, {Goobar}, {Groom}, {Hook}, {Kashikawa}, {Kim}, {Knop}, {Lee},
  {Mendez}, {Morokuma}, {Motohara}, {Nugent}, {Pain}, {Perlmutter}, {Prasad},
  {Quimby}, {Raux}, {Regnault}, {Ruiz-Lapuente}, {Sainton}, {Schaefer},
  {Schahmaneche}, {Smith}, {Spadafora}, {Stanishev}, {Walton}, {Wang},
  {Wood-Vasey}, \& {Yasuda (The Supernova Cosmology
  Project)}}]{2005A&A...430..843L}
{Lidman}, C., {Howell}, D.~A., {Folatelli}, G., {et~al.} 2005, \aap, 430, 843

\bibitem[{{Matheson} {et~al.}(2005){Matheson}, {Blondin}, {Foley}, {Chornock},
  {Filippenko}, {Leibundgut}, {Smith}, {Sollerman}, {Spyromilio}, {Kirshner},
  {Clocchiatti}, {Aguilera}, {Barris}, {Becker}, {Challis}, {Covarrubias},
  {Garnavich}, {Hicken}, {Jha}, {Krisciunas}, {Li}, {Miceli}, {Miknaitis},
  {Prieto}, {Rest}, {Riess}, {Salvo}, {Schmidt}, {Stubbs}, {Suntzeff}, \&
  {Tonry}}]{2005AJ....129.2352M}
{Matheson}, T., {Blondin}, S., {Foley}, R.~J., {et~al.} 2005, \aj, 129, 2352

\bibitem[{{Mazzali} {et~al.}(2005){Mazzali}, {Benetti}, {Altavilla}, {Blanc},
  {Cappellaro}, {Elias-Rosa}, {Garavini}, {Goobar}, {Harutyunyan}, {Kotak},
  {Leibundgut}, {Lundqvist}, {Mattila}, {Mendez}, {Nobili}, {Pain},
  {Pastorello}, {Patat}, {Pignata}, {Podsiadlowski}, {Ruiz-Lapuente}, {Salvo},
  {Schmidt}, {Sollerman}, {Stanishev}, {Stehle}, {Tout}, {Turatto}, \&
  {Hillebrandt}}]{2005ApJ...623L..37M}
{Mazzali}, P.~A., {Benetti}, S., {Altavilla}, G., {et~al.} 2005, \apjl, 623,
  L37

\bibitem[{{Mazzali} {et~al.}(1997){Mazzali}, {Chugai}, {Turatto}, {Lucy},
  {Danziger}, {Cappellaro}, {della Valle}, \& {Benetti}}]{1997MNRAS.284..151M}
{Mazzali}, P.~A., {Chugai}, N., {Turatto}, M., {et~al.} 1997, \mnras, 284, 151

\bibitem[{{Mazzali} {et~al.}(1993){Mazzali}, {Lucy}, {Danziger}, {Gouiffes},
  {Cappellaro}, \& {Turatto}}]{1993A&A...269..423M}
{Mazzali}, P.~A., {Lucy}, L.~B., {Danziger}, I.~J., {et~al.} 1993, \aap, 269,
  423

\bibitem[{{Meikle} {et~al.}(1996){Meikle}, {Cumming}, {Geballe}, {Lewis},
  {Walton}, {Balcells}, {Cimatti}, {Croom}, {Dhillon}, {Economou}, {Jenkins},
  {Knapen}, {Meadows}, {Morris}, {Perez-Fournon}, {Shanks}, {Smith}, {Tanvir},
  {Veilleux}, {Vilchez}, {Wall}, \& {Lucey}}]{1996MNRAS.281..263M}
{Meikle}, W.~P.~S., {Cumming}, R.~J., {Geballe}, T.~R., {et~al.} 1996, \mnras,
  281, 263

\bibitem[{{Patat} {et~al.}(1996){Patat}, {Benetti}, {Cappellaro}, {Danziger},
  {della Valle}, {Mazzali}, \& {Turatto}}]{1996MNRAS.278..111P}
{Patat}, F., {Benetti}, S., {Cappellaro}, E., {et~al.} 1996, \mnras, 278, 111

\bibitem[{{Perlmutter} {et~al.}(1998){Perlmutter}, {Aldering}, {della Valle},
  {Deustua}, {Ellis}, {Fabbro}, {Fruchter}, {Goldhaber}, {Groom}, {Hook},
  {Kim}, {Kim}, {Knop}, {Lidman}, {McMahon}, {Nugent}, {Pain}, {Panagia},
  {Pennypacker}, {Ruiz-Lapuente}, {Schaefer}, \&
  {Walton}}]{1998Natur.391...51P}
{Perlmutter}, S., {Aldering}, G., {della Valle}, M., {et~al.} 1998, \nat, 391,
  51

\bibitem[{{Perlmutter} {et~al.}(1999){Perlmutter}, {Aldering}, {Goldhaber},
  {Knop}, {Nugent}, {Castro}, {Deustua}, {Fabbro}, {Goobar}, {Groom}, {Hook},
  {Kim}, {Kim}, {Lee}, {Nunes}, {Pain}, {Pennypacker}, {Quimby}, {Lidman},
  {Ellis}, {Irwin}, {McMahon}, {Ruiz-Lapuente}, {Walton}, {Schaefer}, {Boyle},
  {Filippenko}, {Matheson}, {Fruchter}, {Panagia}, {Newberg}, {Couch}, \& {The
  Supernova Cosmology Project}}]{1999ApJ...517..565P}
{Perlmutter}, S., {Aldering}, G., {Goldhaber}, G., {et~al.} 1999, \apj, 517,
  565

\bibitem[{{Phillips} {et~al.}(2002){Phillips}, {Krisciunas}, {Suntzeff},
  {et~al.}}]{Phillips:2002cg}
{Phillips}, M.~M., {Krisciunas}, K., {Suntzeff}, N.~B., {et~al.} 2002, From
  Twilight to Highlight - The Physics of Supernovae, ESO/MPA/MPE Workshop,
  Garching

\bibitem[{{Phillips} {et~al.}(1987){Phillips}, {Phillips}, {Heathcote},
  {Blanco}, {Geisler}, {Hamilton}, {Suntzeff}, {Jablonski}, {Steiner},
  {Cowley}, {Schmidtke}, {Wyckoff}, {Hutchings}, {Tonry}, {Strauss},
  {Thorstensen}, {Honey}, {Maza}, {Ruiz}, {Landolt}, {Uomoto}, {Rich},
  {Grindlay}, {Cohn}, {Smith}, {Lutz}, {Lavery}, \&
  {Saha}}]{1987PASP...99..592P}
{Phillips}, M.~M., {Phillips}, A.~C., {Heathcote}, S.~R., {et~al.} 1987, \pasp,
  99, 592

\bibitem[{{Phillips} {et~al.}(1992){Phillips}, {Wells}, {Suntzeff}, {Hamuy},
  {Leibundgut}, {Kirshner}, \& {Foltz}}]{1992AJ....103.1632P}
{Phillips}, M.~M., {Wells}, L.~A., {Suntzeff}, N.~B., {et~al.} 1992, \aj, 103,
  1632

\bibitem[{{Riess} {et~al.}(1998){Riess}, {Filippenko}, {Challis},
  {Clocchiatti}, {Diercks}, {Garnavich}, {Gilliland}, {Hogan}, {Jha},
  {Kirshner}, {Leibundgut}, {Phillips}, {Reiss}, {Schmidt}, {Schommer},
  {Smith}, {Spyromilio}, {Stubbs}, {Suntzeff}, \&
  {Tonry}}]{1998AJ....116.1009R}
{Riess}, A.~G., {Filippenko}, A.~V., {Challis}, P., {et~al.} 1998, \aj, 116,
  1009

\bibitem[{{Riess} {et~al.}(2003){Riess}, {Strolger}, {Tonry}, {Tsvetanov},
  {Casertano}, {Ferguson}, {Mobasher}, {Challis}, {Panagia}, {Filippenko},
  {Li}, {Chornock}, {Kirshner}, {Leibundgut}, {Dickinson}, {Koekemoer},
  {Grogin}, \& {Giavalisco}}]{2003astro.ph..8185R}
{Riess}, A.~G., {Strolger}, L., {Tonry}, J., {et~al.} 2003, astro-ph/0308185

\bibitem[{{Riess} {et~al.}(2004){Riess}, {Strolger}, {Tonry}, {Casertano},
  {Ferguson}, {Mobasher}, {Challis}, {Filippenko}, {Jha}, {Li}, {Chornock},
  {Kirshner}, {Leibundgut}, {Dickinson}, {Livio}, {Giavalisco}, {Steidel},
  {Ben{\'{\i}}tez}, \& {Tsvetanov}}]{2004ApJ...607..665R}
{Riess}, A.~G., {Strolger}, L.-G., {Tonry}, J., {et~al.} 2004, \apj, 607, 665

\bibitem[{{Ruiz-Lapuente} {et~al.}(1992){Ruiz-Lapuente}, {Cappellaro},
  {Turatto}, {Gouiffes}, {Danziger}, {della Valle}, \&
  {Lucy}}]{1992ApJ...387L..33R}
{Ruiz-Lapuente}, P., {Cappellaro}, E., {Turatto}, M., {et~al.} 1992, \apjl,
  387, L33

\bibitem[{{Schaefer} {et~al.}(1999){Schaefer}, {Snyder}, {Hernandez},
  {Roscherr}, {Deng}, {Ellman}, {Bailyn}, {Rengstorf}, {Smith}, {Levine},
  {Barthelmy}, {Butterworth}, {Hurley}, {Cline}, {Meegan}, {Kouveliotou},
  {Kippen}, {Park}, {Williams}, {Porrata}, {Bionta}, {Hartmann}, {Band},
  {Frail}, {Kulkarni}, {Bloom}, {Djorgovski}, {Sadava}, {Chaffee}, {Harris},
  {Abad}, {Adams}, {Andrews}, {Baltay}, {Bongiovanni}, {Briceno}, {Bruzual},
  {Coppi}, {della Prugna}, {Dubuc}, {Emmet}, {Ferrin}, {Fuenmayor}, {Gebhard},
  {Herrera}, {Honeycutt}, {Magris}, {Mateu}, {Muffson}, {Musser}, {Naranjo},
  {Oemler}, {Pacheco}, {Paredes}, {Rengel}, {Romero}, {Rosenzweig}, {Sabbey},
  {S{\'a}nchez}, {S{\'a}nchez}, {Schenner}, {Shin}, {Sinnott}, {Sofia},
  {Stock}, {Suarez}, {Tell{\'e}ria}, {Vicente}, {Vieira}, \&
  {Vivas}}]{1999ApJ...524L.103S}
{Schaefer}, B.~E., {Snyder}, J.~A., {Hernandez}, J., {et~al.} 1999, \apjl, 524,
  L103

\bibitem[{{Schlegel} {et~al.}(1998){Schlegel}, {Finkbeiner}, \&
  {Davis}}]{1998ApJ...500..525S}
{Schlegel}, D.~J., {Finkbeiner}, D.~P., \& {Davis}, M. 1998, \apj, 500, 525

\bibitem[{{Schmidt} {et~al.}(1998){Schmidt}, {Suntzeff}, {Phillips},
  {Schommer}, {Clocchiatti}, {Kirshner}, {Garnavich}, {Challis}, {Leibundgut},
  {Spyromilio}, {Riess}, {Filippenko}, {Hamuy}, {Smith}, {Hogan}, {Stubbs},
  {Diercks}, {Reiss}, {Gilliland}, {Tonry}, {Maza}, {Dressler}, {Walsh}, \&
  {Ciardullo}}]{1998ApJ...507...46S}
{Schmidt}, B.~P., {Suntzeff}, N.~B., {Phillips}, M.~M., {et~al.} 1998, \apj,
  507, 46

\bibitem[{{Strolger} {et~al.}(2002){Strolger}, {Smith}, {Suntzeff}, {Phillips},
  {Aldering}, {Nugent}, {Knop}, {Perlmutter}, {Schommer}, {Ho}, {Hamuy},
  {Krisciunas}, {Germany}, {Covarrubias}, {Candia}, {Athey}, {Blanc},
  {Bonacic}, {Bowers}, {Conley}, {Dahl{\'e}n}, {Freedman}, {Galaz}, {Gates},
  {Goldhaber}, {Goobar}, {Groom}, {Hook}, {Marzke}, {Mateo}, {McCarthy},
  {M{\'e}ndez}, {Muena}, {Persson}, {Quimby}, {Roth}, {Ruiz-Lapuente},
  {Seguel}, {Szentgyorgyi}, {von Braun}, {Wood-Vasey}, \&
  {York}}]{2002AJ....124.2905S}
{Strolger}, L.-G., {Smith}, R.~C., {Suntzeff}, N.~B., {et~al.} 2002, \aj, 124,
  2905

\bibitem[{{Tonry} {et~al.}(2003){Tonry}, {Schmidt}, {Barris}, {Candia},
  {Challis}, {Clocchiatti}, {Coil}, {Filippenko}, {Garnavich}, {Hogan},
  {Holland}, {Jha}, {Kirshner}, {Krisciunas}, {Leibundgut}, {Li}, {Matheson},
  {Phillips}, {Riess}, {Schommer}, {Smith}, {Sollerman}, {Spyromilio},
  {Stubbs}, \& {Suntzeff}}]{2003ApJ...594....1T}
{Tonry}, J.~L., {Schmidt}, B.~P., {Barris}, B., {et~al.} 2003, \apj, 594, 1

\bibitem[{{Vink{\'o}} {et~al.}(2001){Vink{\'o}}, {Kiss}, {Cs{\'a}k}, {F{\H
  u}r{\'e}sz}, {Szab{\'o}}, {Thomson}, \& {Mochnacki}}]{2001AJ....121.3127V}
{Vink{\'o}}, J., {Kiss}, L.~L., {Cs{\'a}k}, B., {et~al.} 2001, \aj, 121, 3127

\bibitem[{{Wells} {et~al.}(1994){Wells}, {Phillips}, {Suntzeff}, {Heathcote},
  {Hamuy}, {Navarrete}, {Fernandez}, {Weller}, {Schommer}, {Kirshner},
  {Leibundgut}, {Willner}, {Peletier}, {Schlegel}, {Wheeler}, {Harkness},
  {Bell}, {Matthews}, {Filippenko}, {Shields}, {Richmond}, {Jewitt}, {Luu},
  {Tran}, {Appleton}, {Robson}, {Tyson}, {Guhathakurta}, {Eder}, {Bond},
  {Potter}, {Veilleux}, {Porter}, {Humphreys}, {Janes}, {Williams}, {Costa},
  {Ruiz}, {Lee}, {Lutz}, {Rich}, {Winkler}, \& {Tyson}}]{1994AJ....108.2233W}
{Wells}, L.~A., {Phillips}, M.~M., {Suntzeff}, B., {et~al.} 1994, \aj, 108,
  2233

\bibitem[{{Wood-Vasey} {et~al.}(2007){Wood-Vasey}, {Miknaitis}, {Stubbs},
  {Jha}, {Riess}, {Garnavich}, {Kirshner}, {Aguilera}, {Becker}, {Blackman},
  {Blondin}, {Challis}, {Clocchiatti}, {Conley}, {Covarrubias}, {Davis},
  {Filippenko}, {Foley}, {Garg}, {Hicken}, {Krisciunas}, {Leibundgut}, {Li},
  {Matheson}, {Miceli}, {Narayan}, {Pignata}, {Prieto}, {Rest}, {Salvo},
  {Schmidt}, {Smith}, {Sollerman}, {Spyromilio}, {Tonry}, {Suntzeff}, \&
  {Zenteno}}]{2007astro.ph..1041W}
{Wood-Vasey}, W.~M., {Miknaitis}, G., {Stubbs}, C.~W., {et~al.} 2007, ArXiv
  Astrophysics e-prints

\end{thebibliography}
\end{document}